\documentclass[fleqn,onecolumn,showpacs,preprintnumbers,showkeys]{revtex4}
\usepackage{graphicx}
\usepackage{amssymb}
\usepackage{amsmath}
\usepackage{bm}
\usepackage[flushleft]{caption2}

\begin{document}
\setcounter{page}{1}
\title{Magneto-optic and electro-optic effects in electromagnetic and gravitational fields}
\author{Zi-Hua Weng}
\email{xmuwzh@xmu.edu.cn.}
\affiliation{School of Physics and Mechanical \& Electrical Engineering,
\\Xiamen University, Xiamen 361005, China}

\begin{abstract}
The magneto-optic effects and electro-optic effects are the essential optic effects, although their theoretical explanations are not unified in the classical electromagnetic theory. Describing with the algebra of octonions, the electromagnetic theory can derive the magneto-optic effects and electro-optic effects from the same one force definition in the paper. As well as the magneto-optic effect is deduced from the known force term, the electro-optic effect is from the new force term in the octonion space. This description method is different to that in the quantum theory as well as the index ellipsoid approach for the electro-optic effect. One more significant inference is that the gravitational field has an impact on the rotation of linearly polarized light, and then results in the similar birefringence for the transmitted light in the cosmic matter.
\end{abstract}



\maketitle

\section{Introduction}

The magneto-optic effects and electro-optic effects can be used to explain the birefringent phenomenons and the rotation of polarized plane in the optical medium. The magneto-optic effect includes the Larmor precession \cite{larmor, fitzgerald}, Zeeman effect \cite{zeeman}, and Faraday effect \cite{faraday} in the magnetic field. The Larmor precession is the precession of the magnetic moments of charged particles in the external magnetic field. The Zeeman effect is the splitting of a spectral line into several components in the presence of a static magnetic field, while the Faraday effect is an interaction between the light and magnetic field in a medium. The electro-optic effect involves the Stark effect \cite{stark}, Pockels effect \cite{pockels}, and Kerr effect \cite{kerr} in the electric field. The Stark effect is the electric analogue of the Zeeman effect, and is the shifting and splitting of spectral lines of charged particles in the external static electric field. The Pockels effect originates the birefringence in an optical medium, and the birefringence is proportional to the electric field, whereas in the Kerr effect it is quadratic in the field.

At present two theoretical methods have been proposed to explain the above optic effects. The theoretical basis of the magneto-optic effects was developed by J. MacCullagh \cite{maccullagh}, G. B. Airy, J. C. Maxwell \cite{maxwell}, and H. A. Rowlland \cite{rowlland1} etc. While the refringent index ellipsoid approach \cite{barrett} was presented to describe the electro-optic effects.

In the magnetic field, when the magnetic moment is placed in the magnetic field, it will tend to align with the magnetic field, that is the magnetic field exerts a torque on the magnetic moment. The Larmor precession can explain the Zeeman effect. The latter is important in the applications, including the nuclear magnetic resonance spectroscopy, the electron spin resonance spectroscopy, and the magnetic resonance imaging etc. The Zeeman effect encompasses the inverse Zeeman effect, anomalous Zeeman effect, and Paschen-Back effect \cite{nawaz} etc. And it can be used to understand qualitatively the circular birefringence in the Faraday effect, in which the rotation of polarized plane is proportional to the intensity component of the applied magnetic field in the direction of light beam. The Faraday effect corroborates that the light and electromagnetism are related. Its theoretical explanation is that light waves is decomposed into two circularly polarized rays with different speeds in the medium, and then the difference of speeds will cause a net phase offset and result in a rotation angle of linear polarization.

In the electric field, the Stark effect includes its first-order, second-order, and inverse effects. The first-order Stark effect is linear in the applied electric field, while the second-order Stark effect is quadratic in the field. The Pockels effect can explain some birefringent phenomenons in the crystals and noncentrosymmetric media, including the lithium niobate, gallium arsenide, electric-field poled polymers, and glasses etc. Meanwhile the Kerr effect may unpuzzle the some double refractions in the solid and liquid dielectrics, when the change of refractive index is directly proportional to the square of the electric field. All materials show a Kerr effect, but certain liquids display it more strongly than others. However it is evident that there is not the electric analogue of the Larmor precession in the electric field. And the Stark effect seems to be independent of the Pockels effect as well as the Kerr effect theoretically. This situation is quite distinct to that in the magneto-optic effects, and it is an unsymmetrical case.

In the electromagnetic and gravitational fields described with the algebra of octonion \cite{cayley}, the magneto-optic effects, electro-optic effects, and some new optic effects can be deduced from the same force definition in the octonion space. Especially the electro-optic effects can be derived from the new force terms, including the Stark effect, Pockels effect, and Kerr effect etc. One more interesting result is that the gravitational field has also an influence on the orientational rotation of linearly polarized light in the long-distance transmission.

\section{Force in electromagnetic and gravitational fields}

The algebra of quaternions \cite{hamilton} was first used by J. C. Maxwell to describe the electromagnetic theory. At present the gravitational theory can be depicted with the quaternion also. But the quaternion space for gravitational field is independent to that for electromagnetic field \cite{weng1}, and these two quaternion spaces are perpendicular to each other, according to the 'Spacetime equality postulation' \cite{weng2}. Two quaternion spaces can be combined together to become one octonion space, consequently the properties of gravitational field and electromagnetic field can be described with the algebra of octonions at the same time.

The force terms of electromagnetic and gravitational fields can be defined from the torque and the energy in the octonion space. This force definition covers all known force terms in the classical theories, including the inertial force, Newtonian gravitational force \cite{newton}, Coulomb electric force, Lorentz magnetic force \cite{lorentz}, gradient of energy, and interacting force between the magnetic strength with magnetic moment, etc.

\subsection{Linear momentum}

In the octonion space, we can define the linear momentum. In the quaternion space for the gravitational field, the basis vector is $\mathbb{E}_g$ = ($1$, $\emph{\textbf{i}}_1$, $\emph{\textbf{i}}_2$, $\emph{\textbf{i}}_3$), the radius vector is $\mathbb{R}_g$ = ($r_0$, $r_1$, $r_2$, $r_3$), and the velocity is $\mathbb{V}_g$ = ($v_0$, $v_1$, $v_2$, $v_3$). In that for the electromagnetic field, the basis vector is $\mathbb{E}_e$ = ($\emph{\textbf{I}}_0$, $\emph{\textbf{I}}_1$, $\emph{\textbf{I}}_2$, $\emph{\textbf{I}}_3$), the radius vector is $\mathbb{R}_e$ = ($R_0$, $R_1$, $R_2$, $R_3$), and the velocity is $\mathbb{V}_e$ = ($V_0$, $V_1$, $V_2$, $V_3$), with $\mathbb{E}_e$ = $\mathbb{E}_g$ $\circ$ $\emph{\textbf{I}}_0$ . The $\mathbb{E}_e$ is independent of the $\mathbb{E}_g$ , and that they can combine together to become the basis vector $\mathbb{E}$ of octonion space, that is $ \mathbb{E} = (1, \emph{\textbf{i}}_1, \emph{\textbf{i}}_2, \emph{\textbf{i}}_3, \emph{\textbf{I}}_0, \emph{\textbf{I}}_1, \emph{\textbf{I}}_2, \emph{\textbf{I}}_3) $ .

The radius vector $\mathbb{R}$ in the octonion space is
\begin{eqnarray}
\mathbb{R} = r_0 + \emph{\textbf{i}}_1 r_1 + \emph{\textbf{i}}_2
r_2 + \emph{\textbf{i}}_3 r_3 + k_{eg} ( \emph{\textbf{I}}_0 R_0 + \emph{\textbf{I}}_1 R_1 +
\emph{\textbf{I}}_2 R_2 + \emph{\textbf{I}}_3 R_3)~,
\end{eqnarray}
and the velocity $\mathbb{V}$ is
\begin{eqnarray}
\mathbb{V} = v_0 + \emph{\textbf{i}}_1 v_1 + \emph{\textbf{i}}_2
v_2 + \emph{\textbf{i}}_3 v_3 +  k_{eg} ( \emph{\textbf{I}}_0 V_0 + \emph{\textbf{I}}_1 V_1 +
\emph{\textbf{I}}_2 V_2 + \emph{\textbf{I}}_3 V_3)~,
\end{eqnarray}
where $r_0 = v_0 t$, $t$ is the time; $v_0$ is the speed of gravitational intermediate boson, $V_0$ is the speed of electromagnetic intermediate boson; the symbol $\circ$ denotes the octonion multiplication.

The gravitational potential $\mathbb{A}_g = (a_0 , a_1 , a_2 , a_3)$ combines with electromagnetic potential $\mathbb{A}_e = (A_0 , A_1 , A_2 , A_3)$ to become the octonion field potential $\mathbb{A} = \mathbb{A}_g + k_{eg} \mathbb{A}_e $ . While the field strength $\mathbb{B} = \Sigma ( b_i \emph{\textbf{i}}_i + k_{eg} B_i \emph{\textbf{I}}_i)$ consists of gravitational strength $\mathbb{B}_g$ and electromagnetic strength $\mathbb{B}_e$ , with the gauge $b_0 = 0$ and $B_0 = 0$ ,
\begin{eqnarray}
\mathbb{B} = \lozenge \circ \mathbb{A} = \mathbb{B}_g + k_{eg} \mathbb{B}_e~,
\end{eqnarray}
where $k_{eg}$ is one coefficient for the dimensional homogeneity. $\lozenge = \Sigma (\emph{\textbf{i}}_i \partial_i)$, with $\partial_i = \partial / \partial r_i$. $i = 0, 1, 2, 3$.

The gravitational strength $\mathbb{B}_g$ includes two components, $\textbf{g} = ( g_{01} , g_{02} , g_{03} ) $ and $\textbf{b} = ( g_{23} , g_{31} , g_{12} )$,
\begin{eqnarray}
\textbf{g}/v_0 = \emph{\textbf{i}}_1 ( \partial_0 a_1 +
\partial_1 a_0 ) + \emph{\textbf{i}}_2 ( \partial_0 a_2 + \partial_2
a_0 ) + \emph{\textbf{i}}_3 ( \partial_0 a_3 + \partial_3 a_0 )~,
\\
\textbf{b} = \emph{\textbf{i}}_1 ( \partial_2 a_3 -
\partial_3 a_2 ) + \emph{\textbf{i}}_2 ( \partial_3 a_1 - \partial_1
a_3 ) + \emph{\textbf{i}}_3 ( \partial_1 a_2 - \partial_2 a_1 )~,
\end{eqnarray}
meanwhile the electromagnetic strength $\mathbb{B}_e$ involves two parts, $\textbf{E} = ( B_{01} , B_{02} , B_{03} ) $ and $\textbf{B} = ( B_{23} , B_{31} , B_{12} )$ ,
\begin{eqnarray}
\textbf{E}/v_0 = \emph{\textbf{I}}_1 ( \partial_0 A_1 +
\partial_1 A_0 ) + \emph{\textbf{I}}_2 ( \partial_0 A_2 + \partial_2
A_0 ) + \emph{\textbf{I}}_3 ( \partial_0 A_3 + \partial_3 A_0 )~,
\\
\textbf{B} = \emph{\textbf{I}}_1 ( \partial_3 A_2 - \partial_2
A_3 ) + \emph{\textbf{I}}_2 ( \partial_1 A_3 - \partial_3 A_1 )
 + \emph{\textbf{I}}_3 ( \partial_2 A_1 - \partial_1 A_2 )~.
\end{eqnarray}

\begin{table}[b]
\caption{\label{tab:table1}The octonion multiplication table.}
\begin{ruledtabular}
\begin{tabular}{ccccccccc}
$ $ & $1$ & $\emph{\textbf{i}}_1$  & $\emph{\textbf{i}}_2$ &
$\emph{\textbf{i}}_3$  & $\emph{\textbf{I}}_0$  &
$\emph{\textbf{I}}_1$
& $\emph{\textbf{I}}_2$  & $\emph{\textbf{I}}_3$  \\
\hline $1$ & $1$ & $\emph{\textbf{i}}_1$  & $\emph{\textbf{i}}_2$ &
$\emph{\textbf{i}}_3$  & $\emph{\textbf{I}}_0$  &
$\emph{\textbf{I}}_1$
& $\emph{\textbf{I}}_2$  & $\emph{\textbf{I}}_3$  \\
$\emph{\textbf{i}}_1$ & $\emph{\textbf{i}}_1$ & $-1$ &
$\emph{\textbf{i}}_3$  & $-\emph{\textbf{i}}_2$ &
$\emph{\textbf{I}}_1$
& $-\emph{\textbf{I}}_0$ & $-\emph{\textbf{I}}_3$ & $\emph{\textbf{I}}_2$  \\
$\emph{\textbf{i}}_2$ & $\emph{\textbf{i}}_2$ &
$-\emph{\textbf{i}}_3$ & $-1$ & $\emph{\textbf{i}}_1$  &
$\emph{\textbf{I}}_2$  & $\emph{\textbf{I}}_3$
& $-\emph{\textbf{I}}_0$ & $-\emph{\textbf{I}}_1$ \\
$\emph{\textbf{i}}_3$ & $\emph{\textbf{i}}_3$ &
$\emph{\textbf{i}}_2$ & $-\emph{\textbf{i}}_1$ & $-1$ &
$\emph{\textbf{I}}_3$  & $-\emph{\textbf{I}}_2$
& $\emph{\textbf{I}}_1$  & $-\emph{\textbf{I}}_0$ \\
\hline $\emph{\textbf{I}}_0$ & $\emph{\textbf{I}}_0$ &
$-\emph{\textbf{I}}_1$ & $-\emph{\textbf{I}}_2$ &
$-\emph{\textbf{I}}_3$ & $-1$ & $\emph{\textbf{i}}_1$
& $\emph{\textbf{i}}_2$  & $\emph{\textbf{i}}_3$  \\
$\emph{\textbf{I}}_1$ & $\emph{\textbf{I}}_1$ &
$\emph{\textbf{I}}_0$ & $-\emph{\textbf{I}}_3$ &
$\emph{\textbf{I}}_2$  & $-\emph{\textbf{i}}_1$
& $-1$ & $-\emph{\textbf{i}}_3$ & $\emph{\textbf{i}}_2$  \\
$\emph{\textbf{I}}_2$ & $\emph{\textbf{I}}_2$ &
$\emph{\textbf{I}}_3$ & $\emph{\textbf{I}}_0$  &
$-\emph{\textbf{I}}_1$ & $-\emph{\textbf{i}}_2$
& $\emph{\textbf{i}}_3$  & $-1$ & $-\emph{\textbf{i}}_1$ \\
$\emph{\textbf{I}}_3$ & $\emph{\textbf{I}}_3$ &
$-\emph{\textbf{I}}_2$ & $\emph{\textbf{I}}_1$  &
$\emph{\textbf{I}}_0$  & $-\emph{\textbf{i}}_3$
& $-\emph{\textbf{i}}_2$ & $\emph{\textbf{i}}_1$  & $-1$ \\
\end{tabular}
\end{ruledtabular}
\end{table}

In the octonion space, the linear momentum density $\mathbb{S}_g$ is the source for the gravitational field, while the electric current density $\mathbb{S}_e$ is that for the electromagnetic field. The octonion field source $\mathbb{S}$ satisfies
\begin{eqnarray}
\mu \mathbb{S} = - ( \mathbb{B}/v_0 + \lozenge)^* \circ \mathbb{B} = \mu_g \mathbb{S}_g + k_{eg} \mu_e \mathbb{S}_e - \mathbb{B}^*
\circ \mathbb{B}/v_0~,
\end{eqnarray}
with
\begin{eqnarray}
\mathbb{B}^* \circ \mathbb{B}/ \mu_g = \mathbb{B}_g^* \circ \mathbb{B}_g / \mu_g + \mathbb{B}_e^* \circ \mathbb{B}_e / \mu_e~,
\end{eqnarray}
where $k_{eg}^2 = \mu_g /\mu_e$; $\mu_g$ and $\mu_e$ are the gravitational and electromagnetic constants respectively; $q$ is the electric charge density; $m$ is the mass density; $\mathbb{S}_g = m \mathbb{V}_g $, and $\mathbb{S}_e = q \mathbb{V}_e$; $*$ denotes the conjugate of octonion.

In the case for coexistence of electromagnetic field and gravitational field, the octonion angular momentum density $\mathbb{L}$ is defined from the linear momentum density $\mathbb{P} = \mu \mathbb{S} / \mu_g $ , radius vector $\mathbb{R}$ , and physics quantity $\mathbb{X}$ ,
\begin{eqnarray}
\mathbb{L} = (\mathbb{R} + k_{rx} \mathbb{X} ) \circ \mathbb{P}~,
\end{eqnarray}
where $\mathbb{L} = \Sigma (l_i \emph{\textbf{i}}_i ) + \Sigma (L_i \emph{\textbf{I}}_i)$; $\mathbb{P} = \Sigma (p_i \emph{\textbf{i}}_i ) + \Sigma (P_i \emph{\textbf{I}}_i)$; $\mathbb{X} = \Sigma (x_i \emph{\textbf{i}}_i) + k_{eg} \Sigma (X_i \emph{\textbf{I}}_i)$. $\textbf{l} = \Sigma (l_j \emph{\textbf{i}}_j)$, $\textbf{L}_0 = L_0 \emph{\textbf{I}}_0$, $\textbf{L} = \Sigma (L_j \emph{\textbf{I}}_j)$. $\textbf{P}_0 = P_0 \emph{\textbf{I}}_0 $; $\textbf{P} = \Sigma (P_j \emph{\textbf{I}}_j )$. $\textbf{R}_0 = R_0 \emph{\textbf{I}}_0 $; $\textbf{R} = \Sigma (R_j \emph{\textbf{I}}_j )$. $k_{rx} = 1$ is a coefficient for the dimensional homogeneity. The derivation of octonion quantity $\mathbb{X}$ will yield the gravitational and electromagnetic potentials simultaneously.

\subsection{Energy and torque}

In the case for coexistence of electromagnetic field and gravitational field, the octonion energy density $\mathbb{W}$ is defined from the octonion angular momentum density $\mathbb{L}$ and octonion field strength $\mathbb{B}$ ,
\begin{eqnarray}
\mathbb{W} = v_0 ( \mathbb{B}/v_0 + \lozenge) \circ \mathbb{L}~,
\end{eqnarray}
where $\mathbb{W} = \Sigma (w_i \emph{\textbf{i}}_i ) + \Sigma (W_i \emph{\textbf{I}}_i )$ ; the $-w_0/2$ is the energy density, the $\textbf{w}/2 = \Sigma (w_j \emph{\textbf{i}}_j )/2$ is the torque density.

Expressing the energy density as
\begin{eqnarray}
w_0 = v_0 \partial_0 l_0 + v_0 \nabla \cdot \textbf{l} + (\textbf{g} / v_0 + \textbf{b}) \cdot \textbf{l} + k_{eg} (\textbf{E} / v_0 + \textbf{B}) \cdot \textbf{L}
\end{eqnarray}
where $-w_0/2$ includes the kinetic energy, gravitational potential energy, field energy, work, electric potential energy, magnetic potential energy, the interacting energy between dipole moment with electromagnetic strength, and some new energy terms. $a_0 / v_0 = \partial_0 x_0 + \nabla \cdot \textbf{x}$ and $\textbf{a} = \partial_0 \textbf{x} + \nabla x_0 + \nabla \times \textbf{x}$ are the scalar and vectorial potential of the gravitational field respectively. $\textbf{A}_0 / v_0 = \partial_0 \textbf{X}_0 + \nabla \cdot \textbf{X}$ and $\textbf{A} = \partial_0 \textbf{X} + \nabla \circ \textbf{X}_0 + \nabla \times \textbf{X}$ are the scalar and vectorial potential of the electromagnetic field respectively. $\nabla = \Sigma (\emph{\textbf{i}}_j \partial_j)$, and $j = 1, 2, 3$.

In a similar way, expressing the torque density $\textbf{w}$ as
\begin{eqnarray}
\textbf{w} = v_0 \partial_0 \textbf{l} + v_0 \nabla l_0 + v_0 \nabla \times \textbf{l} + l_0 (\textbf{g} / v_0 + \textbf{b}) + (\textbf{g} / v_0 +
\textbf{b}) \times \textbf{l} + k_{eg} (\textbf{E} / v_0 + \textbf{B}) \times \textbf{L} + k_{eg} (\textbf{E} / v_0 + \textbf{B}) \circ \textbf{L}_0
\end{eqnarray}
where the above includes the torque density caused by the gravity, electromagnetic force, and other force terms etc.

\begin{table}[b]
\caption{\label{tab:table1}The operator and multiplication of the physical quantity in the octonion space.}
\begin{ruledtabular}
\begin{tabular}{ll}
definitions                 &  meanings                                               \\
\hline
$\nabla \cdot \textbf{a}$   &  $-(\partial_1 a_1 + \partial_2 a_2 + \partial_3 a_3)$  \\
$\nabla \times \textbf{a}$  &  $\emph{\textbf{i}}_1 ( \partial_2 a_3
                                 - \partial_3 a_2 ) + \emph{\textbf{i}}_2 ( \partial_3 a_1
                                 - \partial_1 a_3 )
                                 + \emph{\textbf{i}}_3 ( \partial_1 a_2
                                 - \partial_2 a_1 )$                                  \\
$\nabla a_0$                &  $\emph{\textbf{i}}_1 \partial_1 a_0
                                 + \emph{\textbf{i}}_2 \partial_2 a_0
                                 + \emph{\textbf{i}}_3 \partial_3 a_0  $              \\
$\partial_0 \textbf{a}$     &  $\emph{\textbf{i}}_1 \partial_0 a_1
                                 + \emph{\textbf{i}}_2 \partial_0 a_2
                                 + \emph{\textbf{i}}_3 \partial_0 a_3  $              \\
\hline
$\nabla \cdot \textbf{P}$   &  $-(\partial_1 P_1 + \partial_2 P_2 + \partial_3 P_3) \emph{\textbf{I}}_0 $  \\
$\nabla \times \textbf{P}$  &  $-\emph{\textbf{I}}_1 ( \partial_2
                                 P_3 - \partial_3 P_2 ) - \emph{\textbf{I}}_2 ( \partial_3 P_1
                                 - \partial_1 P_3 )
                                 - \emph{\textbf{I}}_3 ( \partial_1 P_2 - \partial_2 P_1 )$    \\
$\nabla \circ \textbf{P}_0$ &  $\emph{\textbf{I}}_1 \partial_1 P_0
                                 + \emph{\textbf{I}}_2 \partial_2 P_0
                                 + \emph{\textbf{I}}_3 \partial_3 P_0  $             \\
$\partial_0 \textbf{P}$     &  $\emph{\textbf{I}}_1 \partial_0 P_1
                                 + \emph{\textbf{I}}_2 \partial_0 P_2
                                 + \emph{\textbf{I}}_3 \partial_0 P_3  $             \\
\end{tabular}
\end{ruledtabular}
\end{table}

\subsection{Power and force}

In the octonion space, the octonion power density $\mathbb{N}$ is defined from the octonion energy density $\mathbb{W}$ ,
\begin{eqnarray}
\mathbb{N} = v_0 ( \mathbb{B}/v_0 + \lozenge)^* \circ \mathbb{W}~,
\end{eqnarray}
where $\mathbb{N} = \Sigma (n_i \emph{\textbf{i}}_i) + \Sigma (N_i \emph{\textbf{I}}_i)$. The $f_0 = - n_0/(2 v_0)$ is the power density, and the  $\textbf{f} = - \textbf{n} / (2 v_0)$ is the force density in the space $\mathbb{E}_g$ , with the vectorial part $\textbf{n} = \Sigma (n_j \emph{\textbf{i}}_j )$. The other two vectorial parts, $\textbf{N}_0 = N_0 \emph{\textbf{I}}_0$ and $\textbf{N} = \Sigma (N_j \emph{\textbf{I}}_j)$, are in the space $\mathbb{E}_e$ , and may not be detected directly at present.

Further expressing the scalar $n_0$ as
\begin{eqnarray}
n_0 = v_0 \partial_0 w_0 + v_0 \nabla^* \cdot \textbf{w} + (\textbf{g}/v_0 + \textbf{b})^* \cdot \textbf{w} + k_{eg} (\textbf{E}/v_0 + \textbf{B})^* \cdot \textbf{W}~.
\end{eqnarray}

In a similar way, expressing the vectorial part $\textbf{n}$ of $\mathbb{N}$ as follows
\begin{eqnarray}
\textbf{n} = && v_0 \nabla^* w_0  + v_0 \partial_0 \textbf{w} + v_0 \nabla^* \times \textbf{w} + (\textbf{g}/v_0 + \textbf{b})^* \times \textbf{w} + w_0 (\textbf{g}/v_0 + \textbf{b})^*
\nonumber\\
&& + k_{eg} (\textbf{E}/v_0 + \textbf{B})^* \times \textbf{W} + k_{eg} (\textbf{E}/v_0 + \textbf{B})^* \circ \textbf{W}_0~.
\end{eqnarray}

The force density $\textbf{f}$ in the gravitational and electromagnetic fields can be written as,
\begin{eqnarray}
- 2 \textbf{f} = && \nabla^* w_0 + \partial_0 \textbf{w} + (\textbf{g}/v_0 + \textbf{b})^* \times \textbf{w}/v_0 + \nabla^* \times \textbf{w} + w_0 (\textbf{g}/v_0 + \textbf{b})^*/v_0
\nonumber\\
&& + k_{eg} (\textbf{E}/v_0 + \textbf{B})^* \times \textbf{W}/v_0 + k_{eg} (\textbf{E}/v_0 + \textbf{B})^* \circ \textbf{W}_0/v_0~,
\end{eqnarray}
where the force density $\textbf{f}$ includes that of the inertial force, gravity, Lorentz force, gradient of energy, and interacting force between
dipole moment with magnetic strength etc. This force definition is much more complicated than that in the classical field theory, and encompasses more new force terms regarding the gradient of energy etc.

\section{Magneto-optic effect}

In the gravitational and electromagnetic fields, the definition of force density can deduce the Larmor precession frequency for charged particles in the applied magnetic field. Further the Larmor precession frequency can be used to explain the Zeeman effect in the applied magnetic field.

\subsection{Zeeman effect for charged particles}

In case the physical system is on the energy steady state, the charged particle's energy density, $w_0$, will be a constant in the magnetic field, that is, $\nabla w_0 = 0$ . When $(\textbf{g}/v_0 + \textbf{b}) = 0$, $x_0 = 0$, $\textbf{E} = 0$, and $\textbf{B}^* \circ \textbf{W}_0/v_0 \approx 0$, Eq.(17) will be reduced to
\begin{eqnarray}
- 2 \textbf{f} \approx && \partial_0 \textbf{w} + \nabla^* \times \textbf{w} + k_{eg} \textbf{B}^* \times \textbf{W}/v_0
\nonumber\\
\approx && \partial_0 \left\{ k_{eg} \textbf{B} \times ( r_0 \textbf{P} ) \right\} + \partial_0 \left\{ v_0 \partial_0 ( r_0 \textbf{p} ) \right\}
+ \nabla^* \times \left\{ v_0 \partial_0 ( \textbf{r} \times \textbf{p} ) + v_0 \textbf{r} \times ( \nabla \times \textbf{p} ) \right\} + k_{eg} \textbf{B}^* \times \textbf{W}/v_0
\nonumber\\
\approx && 2 \textbf{B} \times q \textbf{V} + 2 v_0 \partial_0 \textbf{p} + v_0 \textbf{r} \nabla^* \cdot ( \partial_0 \textbf{p} + \nabla \times \textbf{p} )
~.
\end{eqnarray}

In the octonion coordinate system, for one basis vector $\emph{\textbf{i}}_i$ , the coordinate is $( r_i + R_i \emph{\textbf{I}}_0 )$ . And the octonion linear momentum is $\mathbb{P} = \Sigma  \left\{ \emph{\textbf{i}}_i \circ ( p_i + P_i \emph{\textbf{I}}_0 ) \right\} $ . Let the applied magnetic field $\textbf{B}$ be along the axis $\emph{\textbf{i}}_3$,  the above can be reduced to two component equations when $\textbf{f} = 0$ ,
\begin{eqnarray}
&& \partial_t^2 r_1 + \omega_0^2 r_1 = (-q B / m ) \partial_t r_2~,
\\
&& \partial_t^2 r_2 + \omega_0^2 r_2 = (q B / m ) \partial_t r_1~,
\end{eqnarray}
where $\omega_0^2 = ( v_0 / 2 m ) \nabla^* \cdot ( \partial_0 \textbf{p} + \nabla \times \textbf{p} ) $ , $\textbf{p} = m \partial_t \textbf{r} $  , $\textbf{r} = \emph{\textbf{i}}_1 r_1 + \emph{\textbf{i}}_2 r_2 $ , with $\partial_t = \partial / \partial t$ . The velocity $V \emph{\textbf{I}}_0$ is perpendicular to the axis $\emph{\textbf{i}}_3$ . $B$ is the magnitude of magnetic field $\textbf{B}$ .

Further the solutions of above equations can be written as,
\begin{eqnarray}
r_1 = r_1^0 exp ( i \omega t )~,~r_2 = r_2^0 exp ( i \omega t )~,
\end{eqnarray}
substituting the above in Eqs.(19)$-$(21), we have the result
\begin{eqnarray}
\omega = \omega_0 + (q B / 2m ) \pm (1/2) \left\{ (qB/m)^2 + 4 \omega_0^2 \right\}^{1/2} ~,
\end{eqnarray}
or
\begin{eqnarray}
\omega = \omega_0 - (q B / 2m ) \pm (1/2) \left\{ (qB/m)^2 + 4 \omega_0^2 \right\}^{1/2} ~,
\end{eqnarray}
where $q B / 2m$ is the Larmor precession frequency. $\omega$ is the angular frequency. $r_1^0$ and $r_2^0$ are both the constants.

The Larmor precession of charged particles can be used to explain the splitting of the spectral lines in a magnetic field about Zeeman effect, which was discovered first by Pieter Zeeman in 1896. The present of a magnetic field splits the energy levels and causes the spectral splitting, since the magnetic field tends to align with the magnetic moment and slightly modifies their energies \cite{uhlenbeck}. And that the magnetic field exerts a torque on the magnetic moment.

\subsection{Quadratic Zeeman effect for charged particles}

The first-order Zeeman effect is linear in the applied magnetic field, meanwhile the second-order Zeeman effect is quadratic in the magnetic field. Taking account of other components of force term $k_{eg} \textbf{B}^* \times \textbf{W}/v_0$ , we can deal with the second-order Zeeman effect.

When $(\textbf{g}/v_0 + \textbf{b}) = 0$, $x_0 = 0$, $\textbf{E} = 0$, $\nabla w_0 = 0$, and $\textbf{B}^* \circ \textbf{W}_0/v_0 \approx 0$, Eq.(17) will be reduced to
\begin{eqnarray}
- 2 \textbf{f} \approx && \partial_0 \textbf{w} + \nabla^* \times \textbf{w} + k_{eg} \textbf{B}^* \times \textbf{W}/v_0
\nonumber\\
\approx && \partial_0 \left\{ k_{eg} \textbf{B} \times ( r_0 \textbf{P} ) \right\} + \partial_0 \left\{ v_0 \partial_0 ( r_0 \textbf{p} ) \right\}
+ \nabla^* \times \left\{ v_0 \partial_0 ( \textbf{r} \times \textbf{p} ) + v_0 \textbf{r} \times ( \nabla \times \textbf{p} ) \right\}
+ k_{eg} \textbf{B}^* \times \left\{ k_{eg} \textbf{B} \times \textbf{l} \right\} / v_0
\nonumber\\
\approx && 2 \textbf{B} \times q \textbf{V} + 2 v_0 \partial_0 \textbf{p} + v_0 \textbf{r} \nabla^* \cdot ( \partial_0 \textbf{p} + \nabla \times \textbf{p} ) + ( k_{eg}^2 B^2 / v_0 ) ( \textbf{r} \times \textbf{p} + p_0 \textbf{r} ) ~.
\end{eqnarray}

Setting the applied magnetic field $\textbf{B}$ along the z-axis in the coordinate system $(r_0, r_1, r_2, r_3)$, the above can be separated into two components equations when $\textbf{f} = 0$ ,
\begin{eqnarray}
&& \partial_t^2 r_1 + \omega_0^2 r_1 = (-q B/ m) \partial_t r_2 - k_{eg}^2 B^2 (r_2 \partial_t r_3 - r_3 \partial_t r_2) / v_0~,
\\
&& \partial_t^2 r_2 + \omega_0^2 r_2 = ( q B/ m) \partial_t r_1 + k_{eg}^2 B^2 (r_3 \partial_t r_1 - r_1 \partial_t r_3) / v_0~,
\end{eqnarray}
where $\omega_0^2 = ( v_0 / 2 m ) \nabla^* \cdot ( \partial_0 \textbf{p} + \nabla \times \textbf{p} ) - k_{eg}^2 B^2 / 2$ , $\textbf{p} = m \partial_t \textbf{r} $ , $\textbf{r} = \Sigma ( \emph{\textbf{i}}_j r_j )$ .

Accordingly a series of intricate frequencies, $\omega$, of charged particles can be derived from the above equations in the magnetic field. In considering of more force terms, we may obtain more complicated equations than the above, and find some more elaborate results about the Zeeman effect for charged particles.

It finds out that the Larmor frequencies may be shifted in the energy's unsteady states, $\nabla w_0 \neq 0$. And the Zeeman effects may be influenced by many other force terms in the spacial coordinate system, especially some factors regarding the magnetic field and gravitational field etc.

\subsection{Faraday magneto-optic effect}

In the gravitational field and electromagnetic field, the definition of force density can draw out Rowlland's equation for magneto-optic effect \cite{rowlland2}. Further the Rowlland's equation can be used to explain the Faraday effect in the glass, crystal, and liquid etc.

In case the physical system is on the torque steady state, the torque density of the charged particle, $\textbf{w}$, will be a constant vector in the magnetic field, that is, $(\partial_0 \textbf{w} + \nabla^* \times \textbf{w}) = 0$ . In the case of $(\textbf{g}/v_0 + \textbf{b}) = 0$, $x_0 = 0$, $\textbf{E} = 0$, $\textbf{B}^* \times \textbf{W} / v_0 \approx 0$, and $\textbf{B}^* \circ \textbf{W}_0 / v_0 \approx 0$, Eq.(17) will be reduced to
\begin{eqnarray}
- 2 \textbf{f} \approx && \nabla^* w_0
\nonumber\\
\approx && k_{eg} \nabla^* ( \textbf{A}_0 \circ \textbf{P}_0  + \textbf{A} \cdot \textbf{P} )
\nonumber\\
\approx && k_{eg} ( \nabla^*  \circ \textbf{A}_0 ) \circ \textbf{P}_0 +  k_{eg} \textbf{P} \times ( \nabla^* \times \textbf{A} ) ~.
\end{eqnarray}

When $\textbf{f} = 0$, the above can be reduced further as follows,
\begin{eqnarray}
0 = k_{eg} ( \nabla^* \circ \textbf{A}_0  ) \circ  \textbf{P}_0 +  (\partial_0 \textbf{D} / \mu_e ) \times \textbf{B}
= ( - \textbf{E}/v_0 + \partial_0 \textbf{A} ) \circ q \textbf{V}_0 +  \partial_0 \textbf{D} \times ( \textbf{B}^* / \mu_e )~,
\end{eqnarray}
and then
\begin{eqnarray}
0 = ( - \textbf{E}/v_0 + \partial_0 \textbf{A} ) \circ \emph{\textbf{I}}_0 +  ( 1 / q V_0 ) (\partial_0 \textbf{D} \times \textbf{H}^*) ~,
\end{eqnarray}
or
\begin{eqnarray}
\textbf{E} \circ \emph{\textbf{I}}_0 = \partial_t \textbf{A} \circ \emph{\textbf{I}}_0 + \sigma (\partial_t \textbf{D} \times \textbf{H}^* )  ~,
\end{eqnarray}
where $\textbf{H} = \textbf{B} / \mu_e $ , $\textbf{D} = \varepsilon \textbf{E}$ , $\sigma = ( q V_0 )^{-1} $ . For a tiny volume-distribution region of the optical medium, the $\textbf{H}$, $\textbf{D}$, and $\sigma$ are all mean values.

Further
\begin{eqnarray}
\partial_t \textbf{E} \circ \emph{\textbf{I}}_0 = \partial_t^2 \textbf{A} \circ \emph{\textbf{I}}_0 + \sigma \partial_t (\partial_t \textbf{D} \times \textbf{H}^* )  ~.
\end{eqnarray}

Considering the following relation obtained from Maxwell equation,
\begin{eqnarray}
\partial_t \textbf{E} = v_0^2 \nabla \times \textbf{B} = - v_0^2 \nabla^2 \textbf{A} ~,
\end{eqnarray}
Eq.(31) can be rewritten as
\begin{eqnarray}
v_0^2 \nabla^2 \textbf{A} \circ \emph{\textbf{I}}_0 = \partial_t^2 \textbf{A} \circ \emph{\textbf{I}}_0 + \varepsilon \sigma \partial_t \left\{ (- v_0^2 \nabla^2 \textbf{A}) \times \textbf{H}^* \right\}  ~.
\end{eqnarray}

Setting the linearly polarized light transmit along the z-axis in the coordinate system $(r_0, r_1, r_2, r_3)$, the above can be reduced to Rowlland's equation for magneto-optic effect,
\begin{eqnarray}
&& v_0^2 \partial_3^2 A_1 - \partial_t^2 A_1 + M H_3  \partial_t ( \partial_3^2 A_2 ) = 0~,
\\
&& v_0^2 \partial_3^2 A_2 - \partial_t^2 A_2 - M H_3  \partial_t ( \partial_3^2 A_1 ) = 0~,
\end{eqnarray}
where $H_3$ is the z-axis component of $\textbf{H}$ , with $M = - v_0^2 \varepsilon \sigma$ .

According to the form of above equations, we can suppose that one solution is
\begin{eqnarray}
&&  A_1 = A cos ( \alpha t - \beta r_3 ) cos ( \gamma t )~,
\\
&&  A_2 = A cos ( \alpha t - \beta r_3 ) sin ( \gamma t )~,
\end{eqnarray}
and making the substitution we find
\begin{eqnarray}
&&  0 = \alpha^2 + \gamma^2 - \beta^2 ( v_0^2  + M H_3 \gamma )~,
\\
&&  0 = 2 \gamma - M \beta^2 H_3~,
\end{eqnarray}
and then
\begin{eqnarray}
&& \gamma = M \beta^2 H_3 / 2~,
\\
&& v^2 = ( \alpha / \beta)^2 =  v_0^2  + M H_3 ( 1 - \gamma / 2)~,
\end{eqnarray}
that is,
\begin{eqnarray}
v^2 = v_0^2  + M H_3 ( 1 - M \lambda^2 H_3 / 8 \pi n^2 )~,
\end{eqnarray}
where $v$ and $v_0$ are the speed of light in the medium and in the vacuum respectively. $\alpha$ is the angular frequency of the  transmitted light. $\beta$ is the quantity which is in inverse proportion to the speed of light. $\gamma$ is the rotation angular velocity of linearly polarized light transmitted along the optical axis.

And then the total angle, $\theta$, of rotation of the light beam will evidently be
\begin{eqnarray}
\theta = \gamma D / v = 2 \pi^2 M H_3 n^3 D / v_0 \lambda_0^2~,
\end{eqnarray}
where $n$ is the refractive index, $\lambda$ is the wavelength, and $D$ is the length of substance. $ \beta \lambda_0 = 2 \pi n $ , $ \alpha \lambda_0 = 2 \pi v $ , $ \lambda = \lambda_0 / n $ , and $v = v_0 / n $ . $\lambda_0$ is the wavelength in the vacuum.

Combining with the dispersion formula, $ d \beta / d \alpha = ( n / v ) \left\{ n - \lambda ( d n / d \lambda ) \right\}$ , the rotation angle of linearly polarized light, $\theta$, will be equivalent to the result from Maxwell's magneto-optic equation,
\begin{eqnarray}
\theta = M' (n / \lambda_0 )^2 \left\{ n - \lambda ( d n / d \lambda ) \right\} H_3 D ~,
\end{eqnarray}
where $M'$ is a constant coefficient.

This inference is accordant to the existing results from M. E. Verdet, M. Faraday, J. B. Biot, J. MacCullagh, G. B. Airy, H. A. Rowlland, and J. C. Maxwell etc. When $\textbf{A}$ and $\textbf{H}$ in Eq.(33) are both 3-dimensional quantities, the obtained rotation angle of linearly polarized light will be more complex than the above.

The first-order Faraday effect is linear in the applied magnetic field, while the second-order Faraday effect is quadratic in the magnetic field. Taking account of the contribution from the force term $(\partial_0 \textbf{w} + \nabla^* \times \textbf{w})$ in the force definition Eq.(17), we can deal with the second-order Faraday effect.

\begin{table}[b]
\caption{\label{tab:table1}The optic effects and related equations of electromagnetic field in the octonion space.}
\begin{ruledtabular}
\begin{tabular}{ll}
optic~effects                            &  equations                         \\
\hline
Larmor precession                        &  $0 = \partial_0 \textbf{w} + \nabla^* \times \textbf{w} + k_{eg} \textbf{B}^* \times \textbf{W}/v_0$  \\
Zeeman effect                            &  $0 = \partial_0 \textbf{w} + \nabla^* \times \textbf{w} + k_{eg} \textbf{B}^* \times \textbf{W}/v_0$  \\
Faraday effect                           &  $0 = \nabla^* w_0$ \\
magneto-optic Kerr effect                &  $0 = \nabla^* w_0$ \\
Voigt and Cotton-Mouton effects          &  $0 = \nabla^* w_0 + k_{eg} \textbf{B}^* \times \textbf{W}/v_0$ \\
self-focusing in the magnetic field      &  $0 = \nabla^* w_0$ \\
\hline
electric analogue of Larmor precession   &  $0 = \partial_0 \textbf{w} + \nabla^* \times \textbf{w} + k_{eg} (\textbf{E}/v_0)^* \times \textbf{W}/v_0$ \\
Stark effect                             &  $0 = \partial_0 \textbf{w} + \nabla^* \times \textbf{w} + k_{eg} (\textbf{E}/v_0)^* \times \textbf{W}/v_0$ \\
Pockels effect                           &  $0 = \nabla^* w_0 + \partial_0 \textbf{w} + \nabla^* \times \textbf{w} $  \\
Kerr effect                              &  $0 = \nabla^* w_0 + \partial_0 \textbf{w} + \nabla^* \times \textbf{w} + k_{eg} (\textbf{E}/v_0)^* \times \textbf{W}/v_0$  \\
self-focusing in the electric field      &  $0 = \nabla^* w_0 + \partial_0 \textbf{w} + \nabla^* \times \textbf{w} $  \\
\end{tabular}
\end{ruledtabular}
\end{table}

\subsection{Double refractions and Self-focusing in the magnetic field}

In the Rowlland's equation for the Faraday effect, the transmitted direction of incident ray is paralleled to that of the applied magnetic field. When these two directions are perpendicular to each other, we will obtain the equations which are similar to the Rowlland's equations. And they can be used to explain the Cotton-Mouton effect and the self-focusing phenomena in the magnetic field.

Setting the linearly polarized light transmit along the z-axis in the coordinate system $(r_0, r_1, r_2, r_3)$, and its polarized direction is paralleled to the x-axis, Eq.(33) can be reduced to the equation which is similar to the Rowlland's equation for magneto-optic effect,
\begin{eqnarray}
&& v_0^2 \partial_3^2 A_1 - \partial_t^2 A_1 + M H_2  \partial_t ( \partial_3^2 A_3 ) = 0~,
\\
&& v_0^2 \partial_3^2 A_3 - \partial_t^2 A_3 - M H_2  \partial_t ( \partial_3^2 A_1 ) = 0~,
\end{eqnarray}
where $H_2$ is the y-axis component of $\textbf{H}$ , $M = - v_0^2 \varepsilon \sigma$ .

According to the form of above equations, we can suppose that one solution is
\begin{eqnarray}
&&  A_1 = A cos ( \alpha t - \beta r_3 ) cos ( \gamma t )~,
\\
&&  A_3 = A cos ( \alpha t - \beta r_3 ) sin ( \gamma t )~,
\end{eqnarray}
and making the substitution we find
\begin{eqnarray}
&&  0 = \alpha^2 + \gamma^2 - \beta^2 ( v_0^2  + M H_2 \gamma )~,
\\
&&  0 = 2 \gamma - M \beta^2 H_2~,
\end{eqnarray}
and then
\begin{eqnarray}
&& \gamma = M \beta^2 H_2 / 2~,
\\
&& v^2 = ( \alpha / \beta)^2 =  v_0^2  + M H_2 ( 1 - \gamma / 2)~,
\end{eqnarray}
that is,
\begin{eqnarray}
v^2 = v_0^2  + M H_2 ( 1 - M \lambda^2 H_2 / 8 \pi n^2 )~,
\end{eqnarray}
where $v$ and $v_0$ are the speed of light in the medium and in the vacuum respectively. $\alpha$ is the angular frequency of the  transmitted light. $\beta$ is the quantity which is in inverse proportion to the speed of light. $\gamma$ is the rotation angular velocity of linearly polarized light transmitted along the optical axis.

And then the total angle, $\theta$, of rotation of the light beam will evidently be
\begin{eqnarray}
\theta = \gamma D / v = 2 \pi^2 M H_2 n^3 D / v_0 \lambda_0^2~,
\end{eqnarray}
where $n$ is the refractive index, $\lambda$ is the wavelength, and $D$ is the length of substance. $ \beta \lambda_0 = 2 \pi n $ , $ \alpha \lambda_0 = 2 \pi v $ , $ \lambda = \lambda_0 / n $ , and $v = v_0 / n $ . $\lambda_0$ is the wavelength in the vacuum.

Combining with the dispersion formula, $ d \beta / d \alpha = ( n / v ) \left\{ n - \lambda ( d n / d \lambda ) \right\}$ , the rotation angle of linearly polarized light, $\theta$, can be written as the following result, which is similar to that in the Faraday effect,
\begin{eqnarray}
\theta = M' (n / \lambda_0 )^2 \left\{ n - \lambda ( d n / d \lambda ) \right\} H_2 D ~,
\end{eqnarray}
where $M'$ is a constant coefficient.

The incident ray may possess two polarized components. One is along the x-axis and the locus is the hyperbolic curve with the rotation angle $\theta$ in Eq.(55), while the other along the y-axis and its direction keeps the same. That is the double refractions phenomena in the magnetic field. In the birefringence the polarized component parallel to the applied field keeps its transmitted orientation, while the polarized component perpendicular to the applied field changes its transmitted orientation in the optical medium. When the incident ray is transmitted along the z-axis, there will exist the refracted ray and the double refractions such as the Cotton-Mouton effect. In particular, the applied magnetic field is symmetric with respect to the z-axis, there will be the self-focusing phenomena.

\section{Electro-optic effect}

In the existing physical theories, it was believed that the description of electro-optic effects had to be dealt with the index ellipsoid theory, including that of Stark effect, Pockels effect, and Kerr effect etc. The index ellipsoid equation is derived from the definition formula of electric field energy density, in which the octonion physical quantity can be written as 3$\times$3 matrix. With some new force terms in the octonion space, these electro-optic effects along the optical axis can be deduced from the octonionic gravitational and electromagnetic fields, which is one kind of theoretical explanations, besides the index ellipsoid theory or the quantum theory.

\subsection{Stark effect for charged particles}

Due to the existence of the force term $k_{eg} (\textbf{E}/v_0) \times (r_0 \textbf{P})$ in the definition of torque density $\textbf{w}$ , the electric field intensity $\textbf{E}$ will impact the spectral lines of charged particles just like the magnetic flux density $\textbf{B}$. The inference is consonant with the Stark effect, which was discovered first by Joannes Stark in 1913. The Stark effect is the electric analogue of the Zeeman effect.

In case the physical system is on the energy steady state, the charged particle's energy density, $w_0$, will be a constant in the electric field, that is, $\nabla w_0 = 0$ . When $(\textbf{g}/v_0 + \textbf{b}) = 0$, $\textbf{B} = 0$, $x_0 = 0$, and $\textbf{E}^* \circ \textbf{W}_0 / v_0^2 \approx 0$,  Eq.(17) will be reduced to
\begin{eqnarray}
- 2 \textbf{f} \approx && \partial_0 \textbf{w} + \nabla^* \times \textbf{w} + k_{eg} (\textbf{E}/v_0)^* \times \textbf{W}/v_0
\nonumber\\
\approx && \partial_0 \left\{ k_{eg} (\textbf{E}/v_0 ) \times ( r_0 \textbf{P} ) \right\} + v_0 \partial_0^2 ( r_0 \textbf{p} )
+ \nabla^* \times \left\{ v_0 \partial_0 ( \textbf{r} \times \textbf{p} ) + v_0 \textbf{r} \times ( \nabla \times \textbf{p} ) \right\} + k_{eg} (\textbf{E}/v_0)^* \times \textbf{W}/v_0
\nonumber\\
\approx && 2 (\textbf{E}/v_0 ) \times q \textbf{V} + 2 v_0 \partial_0 \textbf{p} + v_0 \textbf{r} \nabla^* \cdot ( \partial_0 \textbf{p} + \nabla \times \textbf{p} )~,
\end{eqnarray}
where $k_{eg} (\textbf{E}/v_0 ) \times ( r_0 \textbf{P} )$ is one new force term contrasting to the Maxwell's electromagnetic theory, and it causes the Stark effect directly via the Larmor-like precession in the electric field.

Setting the applied electric field $\textbf{E}$ along the z-axis in the coordinate system $(r_0, r_1, r_2, r_3)$, the above can be separated into two component equations when $\textbf{f} = 0$ ,
\begin{eqnarray}
&& \partial_t^2 r_1 + \omega_0^2 r_1 = (-q E / m v_0) \partial_t r_2~,
\\
&& \partial_t^2 r_2 + \omega_0^2 r_2 = ( q E / m v_0) \partial_t r_1~,
\end{eqnarray}
where $\omega_0^2 = v_0 \nabla^* \cdot ( \partial_0 \textbf{p} + \nabla \times \textbf{p} ) / 2$ , $\textbf{p} = m \partial_t \textbf{r} $  , $\textbf{r} = \Sigma ( \emph{\textbf{i}}_j r_j )$ . $E$ is the magnitude of electric strength $\textbf{E}$.

Further the solutions of above equations can be written as follows,
\begin{eqnarray}
r_1 = r_1^0 exp ( i \omega t )~,~r_2 = r_2^0 exp ( i \omega t )~,
\end{eqnarray}
substituting the above in  Eqs.(57) and (58), we have the result
\begin{eqnarray}
\omega = \omega_0 + (q E / 2m v_0 ) \pm (1/2) \left\{ (q E / m v_0)^2 + 4 \omega_0^2 \right\}^{1/2} ~,
\end{eqnarray}
or
\begin{eqnarray}
\omega = \omega_0 - (q E / 2m v_0 ) \pm (1/2) \left\{ (q E / m v_0)^2 + 4 \omega_0^2 \right\}^{1/2} ~,
\end{eqnarray}
where $q E / 2 m v_0$ is the Larmor-like precession frequency in the electric field.

The Larmor-like precession of the charged particles can be used to explain the splitting of the spectral lines in an electric field about the first-order Stark effect. The present of an electric field splits the energy levels and causes the spectral splitting, since the electric field tends to align with the electric dipole moment and slightly modifies their energies. And that the electric field exerts a torque on the electric dipole moment.

\subsection{Quadratic Stark effect for charged particles}

The first-order Stark effect is linear in the applied electric field, while the second-order Stark effect is quadratic in the electric field. Taking account of other components of the force term $k_{eg} (\textbf{E}/v_0)^* \times \textbf{W}/v_0$ , we can deal with the second-order Stark effect.

When $(\textbf{g}/v_0 + \textbf{b}) = 0$, $x_0 = 0$, $\nabla w_0 = 0$, $\textbf{B} = 0$, and $\textbf{E}^* \circ \textbf{W}_0 / v_0^2 \approx 0$, Eq.(17) will be reduced to
\begin{eqnarray}
- 2 \textbf{f} \approx && \partial_0 \textbf{w} + \nabla^* \times \textbf{w} + k_{eg} (\textbf{E}/v_0)^* \times \textbf{W}/v_0
\nonumber\\
\approx && \partial_0 \left\{ k_{eg} (\textbf{E}/v_0 ) \times ( r_0 \textbf{P} ) \right\} + \partial_0 \left\{ v_0 \partial_0 ( r_0 \textbf{p} ) \right\}
+ \nabla^* \times \left\{ v_0 \partial_0 ( \textbf{r} \times \textbf{p} ) + v_0 \textbf{r} \times ( \nabla \times \textbf{p} ) \right\}
\nonumber\\
&& + k_{eg} (\textbf{E}/v_0)^* \times \left\{ k_{eg} (\textbf{E}/v_0) \times \textbf{l} \right\} / v_0
\nonumber\\
\approx && 2 (\textbf{E}/v_0 ) \times q \textbf{V} + 2 v_0 \partial_0 \textbf{p} + v_0 \textbf{r} \nabla^* \cdot ( \partial_0 \textbf{p} + \nabla \times \textbf{p} )
+ ( k_{eg}^2 E^2 /v_0^3 ) ( \textbf{r} \times \textbf{p} + p_0 \textbf{r} )
~.
\end{eqnarray}

Let the applied electric field $\textbf{E}$ be along the z-axis in the coordinate system $(r_0, r_1, r_2, r_3)$, the above can be separated into two components equations when $\textbf{f} = 0$ ,
\begin{eqnarray}
&& \partial_t^2 r_1 + \omega_0^2 r_1 = (-q E / m v_0) \partial_t r_2 - k_{eg}^2 E^2 (r_2 \partial_t r_3 - r_3 \partial_t r_2) / v_0^3~,
\\
&& \partial_t^2 r_2 + \omega_0^2 r_2 = ( q E / m v_0) \partial_t r_1 + k_{eg}^2 E^2 (r_3 \partial_t r_1 - r_1 \partial_t r_3) / v_0^3~,
\end{eqnarray}
where $\omega_0^2 = ( v_0 / 2 m ) \nabla^* \cdot ( \partial_0 \textbf{p} + \nabla \times \textbf{p} ) - k_{eg}^2 E^2 /2 v_0^2$ , $\textbf{p} = m \partial_t \textbf{r} $ , $\textbf{r} = \Sigma ( \emph{\textbf{i}}_j r_j )$ .

Consequently a series of intricate frequencies, $\omega$, of charged particles can be derived from the above equations in the electric field. In considering of more force terms, we may obtain more complicated equations than the above, and find some more elaborate results about the Stark effect for charged particles. It is easy to think out that the Larmor-like frequencies may be shifted in the energy's unsteady states, $\nabla w_0 \neq 0$. The Sark effects will be affected by many force terms, and some factors regarding electric and gravitational fields etc.

\subsection{Pockels electro-optic effect}

In the gravitational field and electromagnetic field, the definition of force density can conclude some equations, which can be used to explain the Pockels electro-optic effect in the glass, crystal, and liquid etc. In 1893 Friedrich Carl Alwin Pockels discovered the electro-optic effect, which can cause the refractive index to vary approximately in proportion to the electric strength.

In case the physical system is on one complex case, and has to be considered the effect of the $\textbf{w}$ besides of the $w_0$. When $(\textbf{g}/v_0 + \textbf{b}) = 0$, $x_0 = 0$, $\textbf{B} = 0$, $\textbf{E}^* \times \textbf{W}_0 / v_0^2 \approx 0$, and $\textbf{E}^* \circ \textbf{W}_0 / v_0^2 \approx 0$, Eq.(17) will be reduced to
\begin{eqnarray}
- 2 \textbf{f} \approx && \nabla^* w_0 + \partial_0 \textbf{w} + \nabla^* \times \textbf{w}
\nonumber\\
\approx && k_{eg} \nabla^* ( \textbf{X}_0 \circ \textbf{Z}_0  + \textbf{X} \cdot \textbf{Z} ) + k_{eg} \partial_0 ( \textbf{X}_0 \circ \textbf{Z}  + \textbf{X} \circ \textbf{Z}_0 ) + k_{eg} \nabla^* \times ( \textbf{X}_0 \circ \textbf{Z}  + \textbf{X} \circ \textbf{Z}_0 )
\nonumber\\
\approx && k_{eg} ( \nabla^*  \circ \textbf{X}_0 ) \circ \textbf{Z}_0 + k_{eg} ( \partial_0 \textbf{X}_0 + \nabla \circ \textbf{X}_0 ) \times \textbf{Z}
~,
\end{eqnarray}
where $\textbf{Z}_0 = \partial_0 \textbf{P}_0 + \nabla \cdot \textbf{P}$, and $\textbf{Z} = \partial_0 \textbf{P} + \nabla \circ \textbf{P}_0 + \nabla \times \textbf{P}$ .

When $\textbf{f} = 0$, the above can be reduced further as follows,
\begin{eqnarray}
0 = ( - \textbf{A}_E / v_0 + \partial_0 \textbf{X} ) \circ  \textbf{Z}_0 + \textbf{A}_E \times \textbf{Z} ~,
\end{eqnarray}
and then
\begin{eqnarray}
0 = ( - \textbf{A}_E / v_0 + \partial_0 \textbf{X} ) \circ \emph{\textbf{I}}_0 + ( 1 / Z_0 ) ( \textbf{A}_E \times \textbf{Z} ) ~, \nonumber
\end{eqnarray}
or
\begin{eqnarray}
\textbf{A}_E \circ \emph{\textbf{I}}_0 = \partial_t \textbf{X} \circ \emph{\textbf{I}}_0 + \sigma ( \textbf{A}_E \times \textbf{Z} ) ~,
\end{eqnarray}
where $\sigma = v_0 / Z_0 $ , $\textbf{A}_E = \partial_0 \textbf{X} + \nabla \circ \textbf{X}_0$ , $\textbf{A}_B = \nabla \times \textbf{X}$ , $\textbf{Z}_0 = Z_0 \emph{\textbf{I}}_0$ . For a tiny volume-distribution of the optical medium, the $\textbf{P}$, $\textbf{A}_E$, $\textbf{A}_B$, and $\sigma$ are all mean values.

Further
\begin{eqnarray}
\partial_t \textbf{A}_E \circ \emph{\textbf{I}}_0 = \partial_t^2 \textbf{X} \circ \emph{\textbf{I}}_0 + \sigma \partial_t ( \textbf{A}_E \times \textbf{Z} )~.
\end{eqnarray}

Considering the following relation obtained from the definition equations, $ \mathbb{A} = \lozenge \circ \mathbb{X} $ and $ \mathbb{X} = \lozenge^* \circ \mathbb{Y} $ ,
\begin{eqnarray}
\partial_t \textbf{A}_E = v_0^2 \nabla \times \textbf{A}_B = - v_0^2 \nabla^2 \textbf{X} ~,
\end{eqnarray}
Eq.(68) can be rewritten as
\begin{eqnarray}
v_0^2 \nabla^2 \textbf{X} \circ \emph{\textbf{I}}_0 = \partial_t^2 \textbf{X} \circ \emph{\textbf{I}}_0 + \sigma \partial_t \left\{ ( - v_0^2 \nabla^2 \textbf{X} ) \times \textbf{Z} \right\}  ~,
\end{eqnarray}
where $\mathbb{Y}$ is an octonion.

In the coordinate system $(r_0, r_1, r_2, r_3)$, setting the linearly polarized light transmit along the z-axis, which is the optical axis in the longitudinal Pockels effect, the above can be reduced to the equation for electro-optic effect,
\begin{eqnarray}
&& v_0^2 \partial_3^2 X_1 - \partial_t^2 X_1 + M Z_3  \partial_t ( \partial_3^2 X_2 ) = 0~,
\\
&& v_0^2 \partial_3^2 X_2 - \partial_t^2 X_2 - M Z_3  \partial_t ( \partial_3^2 X_1 ) = 0~,
\end{eqnarray}
where $Z_3$ is the z-axis component of $\textbf{Z}$ , and $M = - v_0^2 \sigma$ .

According to the form of above equations, we can suppose that one solution is
\begin{eqnarray}
&&  X_1 = X cos ( \alpha t - \beta r_3 ) cos ( \gamma t )~,
\\
&&  X_2 = X cos ( \alpha t - \beta r_3 ) sin ( \gamma t )~,
\end{eqnarray}
and making the substitution we find
\begin{eqnarray}
&&  0 = \alpha^2 + \gamma^2 - \beta^2 ( v_0^2  + M Z_3 \gamma )~,
\\
&&  0 = 2 \gamma - M \beta^2 Z_3~,
\end{eqnarray}
and then
\begin{eqnarray}
&& \gamma = M \beta^2 Z_3 / 2~,
\\
&& v^2 = ( \alpha / \beta)^2 =  v_0^2  + M Z_3 ( 1 - \gamma / 2)~,
\end{eqnarray}
that is,
\begin{eqnarray}
v^2 = v_0^2  + M Z_3 ( 1 - M \lambda^2 Z_3 / 8 \pi n^2 )~,
\end{eqnarray}
where $v$ is the speed of light in the medium. $\beta \lambda = 2 \pi n$ , and $\alpha \lambda = 2 \pi v$ .

And then the total angle, $\theta$, of rotation of the light beam will evidently be
\begin{eqnarray}
\theta = \gamma D / v = 2 \pi^2 M Z_3 n^3 D / v_0 \lambda_0^2~,
\end{eqnarray}
where $n$ is the refractive index, $\lambda$ is the wavelength, and $D$ is the length of substance. $ \beta \lambda_0 = 2 \pi n $ , $ \alpha \lambda_0 = 2 \pi v $ , $ \lambda = \lambda_0 / n $ , and $v = v_0 / n $ . $\lambda_0$ and $v_0$ are the wavelength and the speed of light in the vacuum respectively.

Combining with the dispersion formula, $ d \beta / d \alpha = ( n / v ) \left\{ n - \lambda ( d n / d \lambda ) \right\}$ , the rotation angle of linearly polarized light will be,
\begin{eqnarray}
\theta = M' (n / \lambda_0 )^2 \left\{ n - \lambda ( d n / d \lambda ) \right\} Z_3 D ~,
\end{eqnarray}
where $M'$ is a constant coefficient.

Considering the condition formula $\textbf{Z} = k' \textbf{J}$ and Ohm's law, $\textbf{J} = \sigma' \textbf{E}$ , we obtain the result when $\partial_t E_3 \approx 0 $ ,
\begin{eqnarray}
Z_3 \approx k' ( \partial_t \sigma' ) E_3 ~,
\end{eqnarray}
and then the angle $\theta$ can be written as follows
\begin{eqnarray}
\theta = M'' (n / \lambda_0 )^2 \left\{ n - \lambda ( d n / d \lambda ) \right\} E_3 D ~,
\end{eqnarray}
where $M''$, $k'$, and $\sigma'$ are coefficients. $E_3$ is the z-axis component of $\textbf{E}$ . $\textbf{J}$ is the electric current density.

This inference is accordant to the existing results from F. C. A. Pockels etc. The above can be considered as one kind of theoretical explanations, besides the index ellipsoid theory as well as the quantum theory. And that the angle $\theta$ can be affected by many force terms in the electric field. When $\textbf{X}$ and $\textbf{Z}$ in Eq.(70) are both 3-dimensional quantities, the obtained rotation angle of linearly polarized light will be more complex than the above.

The Pockels effect is linear in the applied electric field, while the Kerr electro-optic effect is quadratic in the electric field. Taking account of the contribution of force terms regarding $k_{eg} (\textbf{E}/v_0)^* \times \textbf{W}$ in Eq.(17) , we may deal with the Kerr electro-optic effect.

\begin{table}[b]
\caption{\label{tab:table1}The comparison of the optic effects in the electric and magnetic fields.}
\begin{ruledtabular}
\begin{tabular}{lll}
magnetic~field                               &  electric~field                             &  description                \\
\hline
Larmor precession                            &  (analogue)                                 &  charged particle           \\
Zeeman effect                                &  Stark effect                               &  charged particle           \\
Faraday effect (first order)                 &  longitudinal Pockels effect                &  parallel orientations      \\
(analogue)                                   &  transversal Pockels effect                 &  perpendicular orientations \\
Faraday effect (second order)                &  longitudinal Kerr effect                   &  parallel orientations      \\
Voigt and Cotton-Mouton effects              &  transversal Kerr effect                    &  perpendicular orientations \\
magneto-optic Kerr effect                    &  (analogue)                                 &  reflected light            \\
self-focusing                                &  self-focusing                              &  axial symmetric            \\
\end{tabular}
\end{ruledtabular}
\end{table}

\subsection{Double refractions and Self-focusing in the electric field}

In the equation for the electro-optic effect, the transmitted direction of incident ray may be paralleled to that of the applied electric field. When these two directions are perpendicular to each other, we will obtain the equations for the double refractions as well as the self-focusing phenomena in the electric field.

Setting the linearly polarized light transmit along the z-axis in the coordinate system $(r_0, r_1, r_2, r_3)$, and its polarized direction is along the x-axis, Eq.(70) can be reduced to the equation which is similar to Pockels effect,
\begin{eqnarray}
&& v_0^2 \partial_3^2 X_1 - \partial_t^2 X_1 + M Z_2  \partial_t ( \partial_3^2 X_3 ) = 0~,
\\
&& v_0^2 \partial_3^2 X_2 - \partial_t^2 X_2 - M Z_2  \partial_t ( \partial_3^2 X_1 ) = 0~,
\end{eqnarray}
where $Z_2$ is the y-axis component of $\textbf{Z}$ , $M = - v_0^2 \sigma$ .

According to the form of above equations, we can suppose that one solution is
\begin{eqnarray}
&&  X_1 = X cos ( \alpha t - \beta r_3 ) cos ( \gamma t )~,
\\
&&  X_3 = X cos ( \alpha t - \beta r_3 ) sin ( \gamma t )~,
\end{eqnarray}
and making the substitution we find
\begin{eqnarray}
&&  0 = \alpha^2 + \gamma^2 - \beta^2 ( v_0^2  + M Z_2 \gamma )~,
\\
&&  0 = 2 \gamma - M \beta^2 Z_2~,
\end{eqnarray}
and then
\begin{eqnarray}
&& \gamma = M \beta^2 Z_2 / 2~,
\\
&& v^2 = ( \alpha / \beta)^2 =  v_0^2  + M Z_2 ( 1 - \gamma / 2)~,
\end{eqnarray}
that is,
\begin{eqnarray}
v^2 = v_0^2  + M Z_2 ( 1 - M \lambda^2 Z_2 / 8 \pi n^2 )~,
\end{eqnarray}
where $v$ is the speed of light in the medium. $\beta \lambda = 2 \pi n$ , and $\alpha \lambda = 2 \pi v$ .

And then the total angle, $\theta$, of rotation of the light beam will evidently be
\begin{eqnarray}
\theta = \gamma D / v = 2 \pi^2 M Z_2 n^3 D / v_0 \lambda_0^2~,
\end{eqnarray}
where $n$ is the refractive index, $\lambda$ is the wavelength, and $D$ is the length of substance. $ \beta \lambda_0 = 2 \pi n $ , $ \alpha \lambda_0 = 2 \pi v $ , $ \lambda = \lambda_0 / n $ , and $v = v_0 / n $ . $\lambda_0$ and $v_0$ are the wavelength and the speed of light in the vacuum respectively.

Combining with the dispersion formula, $ d \beta / d \alpha = ( n / v ) \left\{ n - \lambda ( d n / d \lambda ) \right\}$ , the rotation angle of linearly polarized light can be written as the result which is similar to that in the Pockels effect,
\begin{eqnarray}
\theta = M' (n / \lambda_0 )^2 \left\{ n - \lambda ( d n / d \lambda ) \right\} Z_2 D ~,
\end{eqnarray}
where $M'$ is a constant coefficient.

Considering the condition formula $\textbf{Z} = k' \textbf{J}$ and Ohm's law, $\textbf{J} = \sigma' \textbf{E}$ , we obtain the result when $\partial_t E_2 \approx 0$,
\begin{eqnarray}
Z_2 \approx k' ( \partial_t \sigma' ) E_2 ~,
\end{eqnarray}
and then the angle $\theta$ can be written as
\begin{eqnarray}
\theta = M'' (n / \lambda_0 )^2 \left\{ n - \lambda ( d n / d \lambda ) \right\} E_2 D ~,
\end{eqnarray}
where $M''$, $k'$, and $\sigma'$ are coefficients. $E_2$ is the y-axis component of $\textbf{E}$ . $\textbf{J}$ is the electric current density.

The incident ray may possess two polarized components. One is along the x-axis and the locus is the hyperbolic curve with the rotation angle $\theta$ in Eq.(96), while the other along the y-axis and its direction keeps the same. That is the double refractions phenomena in the electric field. The polarized component parallel to the applied field keeps its transmitted orientation, while the polarized component perpendicular to the applied field changes its transmitted orientation. When the ray is transmitted along the z-axis, there will exist the refracted ray and the double refractions, such as the Pockels effect or Kerr effect. More specifically, the applied electric field is symmetric with respect to the z-axis, the light beams will be converged or diverged spontaneously, and there will be the self-focusing phenomena.

\section{Optic effect for magnetic-like component of gravitational field}

In the octonion space, the gravitational strength possesses two components, which are similar to the electric field intensity and the magnetic flux density respectively \cite{heaviside}. In the gravitational field, the applied magnetic-like component of gravitational strength can cause the rotational mass particles to precess similarly.

The definition of force density can deduce the Larmor-like precession frequency for mass particles in the applied magnetic-like component, $\textbf{b}$, of gravitational field. The Larmor-like precession frequency can be used to explain the Zeeman-like effect in the applied magnetic-like component of gravitational field.

\subsection{Zeeman-like effect for mass particles}

In case the physical system is on the energy steady state, the mass particle's energy density, $w_0$, will be a constant in the magnetic-like component of gravitational field, that is, $\nabla w_0 = 0$ . When $(\textbf{E}/v_0 + \textbf{B}) = 0$, $\textbf{X}_0 = 0$, $\textbf{g} = 0$, and $w_0 \textbf{b}^* / v_0 \approx 0$, Eq.(17) will be reduced to
\begin{eqnarray}
- 2 \textbf{f} \approx && \partial_0 \textbf{w} + \nabla^* \times \textbf{w} + \textbf{b}^* \times \textbf{w}/v_0
\nonumber\\
\approx &&
\partial_0 \left\{ \textbf{b} \times ( r_0 \textbf{p} ) \right\} + \partial_0 \left\{ v_0 \partial_0 ( r_0 \textbf{p} ) \right\}
+ \nabla^* \times \left\{ v_0 \partial_0 ( \textbf{r} \times \textbf{p} ) + v_0 \textbf{r} \times ( \nabla \times \textbf{p} ) \right\} + \textbf{b}^* \times \textbf{w}/v_0
\nonumber\\
\approx && 2 \textbf{b} \times m \textbf{v} + 2 v_0 \partial_0 \textbf{p} + v_0 \textbf{r} \nabla^* \cdot ( \partial_0 \textbf{p} + \nabla \times \textbf{p} )
~.
\end{eqnarray}

Let the applied magnetic-like component of gravitational field, $\textbf{b}$, be along the axis $\emph{\textbf{i}}_3$,  the above can be reduced to three components equations when $\textbf{f} = 0$ ,
\begin{eqnarray}
&& \partial_t^2 r_1 + \omega_0^2 r_1 = - b \partial_t r_2~,
\\
&& \partial_t^2 r_2 + \omega_0^2 r_2 = b \partial_t r_1~,
\\
&& \partial_t^2 r_3 = 0~,
\end{eqnarray}
where $\omega_0^2 = ( v_0 / 2 m ) \nabla^* \cdot ( \partial_0 \textbf{p} + \nabla \times \textbf{p} )$ , $\textbf{p} = m \partial_t \textbf{r} $  , $\textbf{r} = \emph{\textbf{i}}_1 r_1 + \emph{\textbf{i}}_2 r_2 $ . The velocity $\textbf{v}$ is perpendicular to the axis $\emph{\textbf{i}}_3$ . And the $b$ is the magnitude of $\textbf{b}$ , which is corresponded to the angular velocity $\vec{\omega} = \nabla \times \textbf{v}/2$ ,

Further the solutions of above equations can be written as follows,
\begin{eqnarray}
r_1 = r_1^0 exp ( i \omega t )~,~r_2 = r_2^0 exp ( i \omega t )~,
\end{eqnarray}
substituting the above in  Eqs.(98)$-$(100), we have the results,
\begin{eqnarray}
\omega = \omega_0 + ( b / 2 ) \pm (1/2) \left\{ b^2 + 4 \omega_0^2 \right\}^{1/2} ~,
\end{eqnarray}
or
\begin{eqnarray}
\omega = \omega_0 - ( b / 2 ) \pm (1/2) \left\{ b^2 + 4 \omega_0^2 \right\}^{1/2} ~,
\end{eqnarray}
where $ b / 2$ is the Larmor-like precession frequency. $\omega$ is the angular frequency. $r_1^0$ and $r_2^0$ are both constants.

The Larmor-like precession of mass particles can be used to explain the splitting of the spectral lines in a magnetic-like field about Zeeman-like effect. The present of a magnetic-like field splits the energy levels and causes the spectral splitting, since the magnetic-like field tends to align with the magnetic-like moment (the angular momentum) and slightly modifies their energies. And that the magnetic-like field exerts a torque on the magnetic-like moment.

\subsection{Quadratic Zeeman-like effect for mass particles}

The first-order Zeeman-like effect is linear in the applied magnetic-like component of gravitational field, while the second-order Zeeman-like effect is quadratic in the magnetic-like component of gravitational field. Taking account of other parts' contribution of force terms regarding $\textbf{b}$ , we can deal with the second-order Zeeman-like effect.

When $(\textbf{E}/v_0 + \textbf{B}) = 0$, $\textbf{X}_0 = 0$, $\nabla w_0 = 0$, $\textbf{g} = 0$, and $w_0 \textbf{b}^* / v_0 \approx 0$, Eq.(17) will be reduced to
\begin{eqnarray}
- 2 \textbf{f} \approx && \partial_0 \textbf{w} + \nabla^* \times \textbf{w} + \textbf{b}^* \times \textbf{w}/v_0
\nonumber\\
\approx && \partial_0 \left\{ \textbf{b} \times ( r_0 \textbf{p} ) \right\} + \partial_0 \left\{ v_0 \partial_0 ( r_0 \textbf{p} ) \right\}
+ \nabla^* \times \left\{ v_0 \partial_0 ( \textbf{r} \times \textbf{p} ) + v_0 \textbf{r} \times ( \nabla \times \textbf{p} ) \right\}
+ \textbf{b}^* \times ( \textbf{b} \times \textbf{l} ) / v_0
\nonumber\\
\approx && 2 \textbf{b} \times m \textbf{v} + 2 v_0 \partial_0 \textbf{p} + v_0 \textbf{r} \nabla^* \cdot ( \partial_0 \textbf{p} + \nabla \times \textbf{p} ) + ( b^2 / v_0 ) ( \textbf{r} \times \textbf{p} + p_0 \textbf{r} ) ~.
\end{eqnarray}

Setting the applied magnetic-like component of gravitational field $\textbf{b}$ along the z-axis in the coordinate system $(r_0, r_1, r_2, r_3)$, the above can be separated into two components equations when $\textbf{f} = 0$ ,
\begin{eqnarray}
&& \partial_t^2 r_1 + \omega_0^2 r_1 = - b \partial_t r_2 -  b^2 (r_2 \partial_t r_3 - r_3 \partial_t r_2) / v_0~,
\\
&& \partial_t^2 r_2 + \omega_0^2 r_2 = b \partial_t r_1 +  b^2 (r_3 \partial_t r_1 - r_1 \partial_t r_3) / v_0~,
\end{eqnarray}
where $\omega_0^2 = ( v_0 / 2 m ) \nabla^* \cdot ( \partial_0 \textbf{p} + \nabla \times \textbf{p} ) -  b^2 / 2$ , $\textbf{p} = m \partial_t \textbf{r} $ , $\textbf{r} = \Sigma ( \emph{\textbf{i}}_j r_j )$ .

Accordingly a series of intricate frequencies, $\omega$, of mass particles can be derived from the above equations in the magnetic-like field. In considering of more force terms, we may obtain more complicated equations than the above, and find some more elaborate results about the Zeeman-like effect for mass particles with the angular velocity.

It is easy to find out that the Larmor-like frequencies may be shifted in the energy's unsteady states, $\nabla w_0 \neq 0$. The Zeeman-like effects may be influenced by many force terms in the spacial coordinate system, and especially some factors in the magnetic-like component of gravitational field and electromagnetic field etc.

\subsection{Faraday-like effect for mass particles}

On the analogy of the derivation processes of Faraday effect, in the gravitational field and electromagnetic field, the definition of force density can conclude some equations, which can be used to explain Faraday-like effect in the magnetic-like component of gravitational field.

In case the physical system is on the torque steady state, the torque density of the mass particle, $\textbf{w}$, will be a constant vector in the magnetic-like field, that is, $(\partial_0 \textbf{w} + \nabla^* \times \textbf{w}) = 0$ . When $(\textbf{E}/v_0 + \textbf{B}) = 0$, $\textbf{X}_0 = 0$, $\textbf{g} = 0$, $\textbf{b}^* \times \textbf{w} / v_0 \approx 0$, and $w_0 \textbf{b}^* / v_0 \approx 0$, Eq.(17) will be reduced to
\begin{eqnarray}
- 2 \textbf{f} \approx && \nabla^* w_0
\nonumber\\
\approx &&  \nabla^* ( a_0 p_0  + \textbf{a} \cdot \textbf{p} )
\nonumber\\
\approx &&  p_0 \nabla^* a_0 +  \textbf{p} \times ( \nabla^* \times \textbf{a} ) ~.
\end{eqnarray}

When $\textbf{f} = 0$, the above can be reduced further as follows,
\begin{eqnarray}
0 = p_0 \nabla^* a_0 +  ( \varepsilon_g \partial_0 \textbf{g} / \mu_g ) \times \textbf{b}
= m v_0 ( - \textbf{g}/v_0 + \partial_0 \textbf{a} ) +  \varepsilon_g \partial_0 \textbf{g} \times ( \textbf{b}^* / \mu_g )~,
\end{eqnarray}
and then
\begin{eqnarray}
0 = ( - \textbf{g}/v_0 + \partial_0 \textbf{a} )  +  ( \varepsilon_g / m v_0 \mu_g ) \partial_0 \textbf{g} \times \textbf{b}^* ~, \nonumber
\end{eqnarray}
or
\begin{eqnarray}
\textbf{g}/v_0 = \partial_0 \textbf{a}  +  \sigma_g \partial_0 \textbf{g} \times \textbf{b}^* ~,
\end{eqnarray}
where $\sigma_g = \varepsilon_g / (m v_0 \mu_g) $ . For a tiny volume-distribution region of the optical medium, the $\textbf{b}$, $\textbf{g}$, and $\sigma_g$ are all mean values.

Further
\begin{eqnarray}
\partial_t \textbf{g}  = \partial_t^2 \textbf{a} + \sigma_g \partial_t (\partial_t \textbf{g} \times \textbf{b}^* )  ~.
\end{eqnarray}

Considering the following relation obtained from Eq.(8),
\begin{eqnarray}
\partial_t \textbf{g} = v_0^2 \nabla \times \textbf{b} = - v_0^2 \nabla^2 \textbf{a} ~,
\end{eqnarray}
Eq.(110) can be rewritten as
\begin{eqnarray}
v_0^2 \nabla^2 \textbf{a} = \partial_t^2 \textbf{a} + \sigma_g \partial_t \left\{ (- v_0^2 \nabla^2 \textbf{a}) \times \textbf{b}^* \right\}  ~.
\end{eqnarray}

Setting the linearly polarized light transmit along the z-axis in the coordinate system $(r_0, r_1, r_2, r_3)$, the above can be reduced to the optic equation about Faraday-like effect in the magnetic-like component of gravitational field,
\begin{eqnarray}
&& v_0^2 \partial_3^2 a_1 - \partial_t^2 a_1 + M_g b_3  \partial_t ( \partial_3^2 a_2 ) = 0~,
\\
&& v_0^2 \partial_3^2 a_2 - \partial_t^2 a_2 - M_g b_3  \partial_t ( \partial_3^2 a_1 ) = 0~,
\end{eqnarray}
where $b_3$ is the z-axis component of $\textbf{b}$ , and $M_g = - v_0^2 \sigma_g $ .

According to the form of above equations, we can suppose that one solution is
\begin{eqnarray}
&&  a_1 = a cos ( \alpha t - \beta r_3 ) cos ( \gamma t )~,
\\
&&  a_2 = a cos ( \alpha t - \beta r_3 ) sin ( \gamma t )~,
\end{eqnarray}
and making the substitution we find
\begin{eqnarray}
&&  0 = \alpha^2 + \gamma^2 - \beta^2 ( v_0^2  + M_g b_3 \gamma )~,
\\
&&  0 = 2 \gamma - M_g \beta^2 b_3~,
\end{eqnarray}
and then
\begin{eqnarray}
&& \gamma = M_g \beta^2 b_3 / 2~,
\\
&& v^2 = ( \alpha / \beta)^2 =  v_0^2  + M_g b_3 ( 1 - \gamma / 2)~,
\end{eqnarray}
that is,
\begin{eqnarray}
v^2 = v_0^2  + M_g b_3 ( 1 - M_g \lambda^2 b_3 / 8 \pi n^2 )~,
\end{eqnarray}
where $v$ is the speed of light in the medium. $\beta \lambda = 2 \pi n$ , and $\alpha \lambda = 2 \pi v$ .

And then the total angle, $\theta$, of rotation of the light beam will evidently be
\begin{eqnarray}
\theta = \gamma D_g / v = 2 \pi^2 M_g b_3 n^3 D_g / v_0 \lambda_0^2~,
\end{eqnarray}
where $n$ is the refractive index, $\lambda$ is the wavelength, and $D_g$ is the length of substance. $ \beta \lambda_0 = 2 \pi n $ , $ \alpha \lambda_0 = 2 \pi v $ , $ \lambda = \lambda_0 / n $ , and $v = v_0 / n $ . $\lambda_0$ and $v_0$ are the wavelength and the speed of light in the vacuum respectively.

Combining with the dispersion formula, $ d \beta / d \alpha = ( n / v ) \left\{ n - \lambda ( d n / d \lambda ) \right\}$ , the rotation angle of linearly polarized light, $\theta$, will be written as follows,
\begin{eqnarray}
\theta = M_g' (n / \lambda_0 )^2 \left\{ n - \lambda ( d n / d \lambda ) \right\} b_3 D_g ~,
\end{eqnarray}
where $M_g'$ is a constant coefficient.

The above means that in case there were the magnetic-like component of gravitational field, the linearly polarized light will rotate the polarization-plane in the sufficient transmitting region in the gravitational field in the absence of the electromagnetic field. When $\textbf{a}$ and $\textbf{b}$ in Eq.(112) are both 3-dimensional quantities, the obtained rotation angle of linearly polarized light will be more complex than the above.

\subsection{Double refractions and self-focusing in the magnetic-like field}

In the equation for the optic effects in the magnetic-like field, the transmitted orientation of incident ray may be paralleled to that of the applied magnetic-like field. When these two directions are perpendicular to each other, we will obtain the equations for the double refractions as well as the self-focusing phenomena in the magnetic-like field.

Setting the linearly polarized light transmit along the z-axis in the coordinate system $(r_0, r_1, r_2, r_3)$, and its polarized direction is along the x-axis, Eq.(112) can be reduced to the equation which is similar to that for Faraday-like effect,
\begin{eqnarray}
&& v_0^2 \partial_3^2 a_1 - \partial_t^2 a_1 + M_g b_2  \partial_t ( \partial_3^2 a_3 ) = 0~,
\\
&& v_0^2 \partial_3^2 a_3 - \partial_t^2 a_3 - M_g b_2  \partial_t ( \partial_3^2 a_1 ) = 0~,
\end{eqnarray}
where $b_2$ is the y-axis component of $\textbf{b}$ , with $M_g = - v_0^2 \sigma_g $ .

According to the form of above equations, we can suppose that one solution is
\begin{eqnarray}
&&  a_1 = a cos ( \alpha t - \beta r_3 ) cos ( \gamma t )~,
\\
&&  a_3 = a cos ( \alpha t - \beta r_3 ) sin ( \gamma t )~,
\end{eqnarray}
and making the substitution we find
\begin{eqnarray}
&&  0 = \alpha^2 + \gamma^2 - \beta^2 ( v_0^2  + M_g b_2 \gamma )~,
\\
&&  0 = 2 \gamma - M_g \beta^2 b_2~,
\end{eqnarray}
and then
\begin{eqnarray}
&& \gamma = M_g \beta^2 b_2 / 2~,
\\
&& v^2 = ( \alpha / \beta)^2 =  v_0^2  + M_g b_2 ( 1 - \gamma / 2)~,
\end{eqnarray}
that is,
\begin{eqnarray}
v^2 = v_0^2  + M_g b_2 ( 1 - M_g \lambda^2 b_2 / 8 \pi n^2 )~,
\end{eqnarray}
where $v$ is the speed of light in the medium. $\beta \lambda = 2 \pi n$ , and $\alpha \lambda = 2 \pi v$ .

And then the total angle, $\theta$, of rotation of the light beam will evidently be
\begin{eqnarray}
\theta = \gamma D_g / v = 2 \pi^2 M_g b_2 n^3 D_g / v_0 \lambda_0^2~,
\end{eqnarray}
where $n$ is the refractive index, $\lambda$ is the wavelength, and $D_g$ is the length of substance. $ \beta \lambda_0 = 2 \pi n $ , $ \alpha \lambda_0 = 2 \pi v $ , $ \lambda = \lambda_0 / n $ , and $v = v_0 / n $ . $\lambda_0$ and $v_0$ are the wavelength and the speed of light in the vacuum respectively.

Combining with the dispersion formula, $ d \beta / d \alpha = ( n / v ) \left\{ n - \lambda ( d n / d \lambda ) \right\}$ , the rotation angle of linearly polarized light, $\theta$, will be written as follows,
\begin{eqnarray}
\theta = M_g' (n / \lambda_0 )^2 \left\{ n - \lambda ( d n / d \lambda ) \right\} b_2 D_g ~,
\end{eqnarray}
where $M_g'$ is a constant coefficient.

The incident ray may possess two polarized components. One is along the x-axis and the locus is the hyperbolic curve with the rotation angle $\theta$ in Eq.(134), while the other along the y-axis and its direction keeps the same. That is the double refractions phenomena in the magnetic-like field.  The polarized component parallel to the applied field keeps its transmitted orientation, while the polarized component perpendicular to the applied field changes its transmitted orientation in the interstellar space. When the incident ray is transmitted along the z-axis, there will exist the refracted ray and the double refractions, such as the Faraday-like effect. In particular, the applied magnetic-like field is symmetric with respect to the z-axis, there will be the self-focusing phenomena.

\begin{table}[h]
\caption{\label{tab:table1}The analogous optic effects and related equations of gravitational field in the octonion space.}
\begin{ruledtabular}
\begin{tabular}{ll}
optic~effects                                                            &  equations                                            \\
\hline
Larmor-like precession in the magnetic-like field                        &  $0 = \partial_0 \textbf{w} + \nabla^* \times \textbf{w} + \textbf{b}^* \times \textbf{w}/v_0$  \\
Zeeman-like effect                                                       &  $0 = \partial_0 \textbf{w} + \nabla^* \times \textbf{w} + \textbf{b}^* \times \textbf{w}/v_0$  \\
Faraday-like magnetic-like optic effect                                  &  $0 = \nabla^* w_0$     \\
reflected Kerr-like effect in the magnetic-like field                    &  $0 = \nabla^* w_0$     \\
Voigt-like and Cotton-Mouton-like effects in the magnetic-like field     &  $0 = \nabla^* w_0 + \textbf{b}^* \times \textbf{w}/v_0$     \\
self-focusing in the magnetic-like field                                 &  $0 = \nabla^* w_0$     \\
\hline
Larmor-like precession in the electric-like field                        &  $0 = \partial_0 \textbf{w} + \nabla^* \times \textbf{w} + (\textbf{g}/v_0)^* \times \textbf{w}/v_0 $  \\
Stark-like effect                                                        &  $0 = \partial_0 \textbf{w} + \nabla^* \times \textbf{w} + (\textbf{g}/v_0)^* \times \textbf{w}/v_0 $  \\
Pockels-like electric-like optic effect                                  &  $0 = \nabla^* w_0 + \partial_0 \textbf{w} + \nabla^* \times \textbf{w} $  \\
Kerr-like effect in the electric-like field                              &  $0 = \nabla^* w_0 + \partial_0 \textbf{w} + \nabla^* \times \textbf{w} + (\textbf{g}/v_0)^* \times \textbf{w}/v_0$  \\
self-focusing in the electric-like field                                 &  $0 = \nabla^* w_0 + \partial_0 \textbf{w} + \nabla^* \times \textbf{w} $  \\
\end{tabular}
\end{ruledtabular}
\end{table}

\section{Optic effect for electric-like component of gravitational field}

By analogy with Stark effect, Pockels effect, and Kerr effect etc in the electric field, some optic effects for electric-like component of gravitational field can be deduced, including the Stark-like effect and Pockels-like effect etc. With some new force terms in the octonion space, these electro-like optic effects can be derived from the octonion gravitational and electromagnetic fields.

\subsection{Stark-like effect for mass particles}

Due to the existence of the force term $(\textbf{g}/v_0) \times (r_0 \textbf{p})$ in the definition of torque density $\textbf{w}$ , the electric-like component of gravitational field $\textbf{g}$ will impact the spectral lines of the mass particles just like the magnetic-like component of gravitational field $\textbf{b}$. The inference is can be called as the Stark-like effect. The Stark-like effect is the electric-like analogue of the Zeeman-like effect in the gravitational field.

In case the physical system is on the energy steady state, the mass particle's energy density, $w_0$, will be a constant in the electric-like field (or the extended Newtonian gravitational field), that is, $\nabla w_0 = 0$ . When $(\textbf{E}/v_0 + \textbf{B}) = 0$, $\textbf{X}_0 = 0$, $\textbf{b} = 0$, and $w_0 \textbf{g}^* / v_0^2 \approx 0$, Eq.(17) will be reduced to
\begin{eqnarray}
- 2 \textbf{f} \approx && \partial_0 \textbf{w} + \nabla^* \times \textbf{w} + (\textbf{g}/v_0)^* \times \textbf{w}/v_0
\nonumber\\
\approx && \partial_0 \left\{ (\textbf{g}/v_0 ) \times ( r_0 \textbf{p} ) \right\} + \partial_0 \left\{ v_0 \partial_0 ( r_0 \textbf{p} ) \right\}
+ \nabla^* \times \left\{ v_0 \partial_0 ( \textbf{r} \times \textbf{p} ) + v_0 \textbf{r} \times ( \nabla \times \textbf{p} ) \right\} + (\textbf{g}/v_0)^* \times \textbf{w}/v_0
\nonumber\\
\approx && 2 (\textbf{g}/v_0 ) \times m \textbf{v} + 2 v_0 \partial_0 \textbf{p} + v_0 \textbf{r} \nabla^* \cdot ( \partial_0 \textbf{p} + \nabla \times \textbf{p} )~,
\end{eqnarray}
where $(\textbf{g}/v_0 ) \times ( r_0 \textbf{p} )$ is one new force term contrasting to the Newtonian gravitational field, and it causes the Stark-like effect directly by means of the Larmor-like precession in the electric-like field.

Setting the applied electric-like field $\textbf{g}$ along the z-axis in the coordinate system $(r_0, r_1, r_2, r_3)$, the above can be separated into two components equations when $\textbf{f} = 0 $ ,
\begin{eqnarray}
&& \partial_t^2 r_1 + \omega_0^2 r_1 = (- g / v_0) \partial_t r_2~,
\\
&& \partial_t^2 r_2 + \omega_0^2 r_2 = (  g / v_0) \partial_t r_1~,
\end{eqnarray}
where $\omega_0^2 = v_0 \nabla^* \cdot ( \partial_0 \textbf{p} + \nabla \times \textbf{p} ) / 2$ , $\textbf{p} = m \partial_t \textbf{r} $  , $\textbf{r} = \Sigma ( \emph{\textbf{i}}_j r_j )$ . $g$ is the magnitude of electric-like field $\textbf{g}$ .

Further the solutions of above equations can be written as follows,
\begin{eqnarray}
r_1 = r_1^0 exp ( i \omega t )~,~r_2 = r_2^0 exp ( i \omega t )~,
\end{eqnarray}
substituting the above in  Eqs.(136) and (137), we have the results
\begin{eqnarray}
\omega = \omega_0 + ( g / 2 v_0 ) \pm (1/2) \left\{ ( g / v_0)^2 + 4 \omega_0^2 \right\}^{1/2} ~,
\end{eqnarray}
or
\begin{eqnarray}
\omega = \omega_0 - ( g / 2 v_0 ) \pm (1/2) \left\{ ( g / v_0)^2 + 4 \omega_0^2 \right\}^{1/2} ~,
\end{eqnarray}
where $ g / 2 v_0$ is the Larmor-like precession frequency in the electric-like field.

The Larmor-like precession of mass particles can be used to explain the splitting of the spectral lines in an electric-like field about the first-order Stark-like effect. The present of an electric-like field splits the energy levels and causes the spectral splitting, since the electric-like field tends to align with the electric-like dipole moment (matter-antimatter) and slightly modifies their energies. And the electric-like field exerts a torque on the electric-like dipole moment.

\subsection{Quadratic Stark-like effect for mass particles}

The first-order Stark-like effect is linear in the applied electric-like field, while the second-order Stark-like effect is quadratic in the electric-like field. Taking account of other parts' contribution of force term $(\textbf{g}/v_0)^* \times \textbf{w}/v_0$ , we can deal with the second-order Stark-like effect.

When $ (\textbf{E}/v_0 + \textbf{B}) = 0$, $\textbf{X}_0 = 0$, $\nabla w_0 = 0$, $\textbf{b} = 0$, and $w_0 \textbf{g}^* / v_0^2 \approx 0$, Eq.(17) will be reduced to
\begin{eqnarray}
- 2 \textbf{f} \approx && \partial_0 \textbf{w} + \nabla^* \times \textbf{w} + (\textbf{g}/v_0)^* \times \textbf{w}/v_0
\nonumber\\
\approx && \partial_0 \left\{ (\textbf{g}/v_0 ) \times ( r_0 \textbf{p} ) \right\} + \partial_0 \left\{ v_0 \partial_0 ( r_0 \textbf{p} ) \right\}
+ \nabla^* \times \left\{ v_0 \partial_0 ( \textbf{r} \times \textbf{p} ) + v_0 \textbf{r} \times ( \nabla \times \textbf{p} ) \right\}
+ \textbf{g}^* \times (\textbf{g} \times \textbf{l} ) / v_0^3
\nonumber\\
\approx && 2 (\textbf{g}/v_0 ) \times m \textbf{v} + 2 v_0 \partial_0 \textbf{p} + v_0 \textbf{r} \nabla^* \cdot ( \partial_0 \textbf{p} + \nabla \times \textbf{p} )
+ ( g^2 /v_0^3 ) ( \textbf{r} \times \textbf{p} + p_0 \textbf{r} )
~.
\end{eqnarray}

Let the applied electric-like field $\textbf{g}$ be along the z-axis in the coordinate system $(r_0, r_1, r_2, r_3)$, the above can be separated into two components equations when $\textbf{f} = 0$ ,
\begin{eqnarray}
&& \partial_t^2 r_1 + \omega_0^2 r_1 = (- g / v_0) \partial_t r_2 - g^2 (r_2 \partial_t r_3 - r_3 \partial_t r_2) / v_0^3~,
\\
&& \partial_t^2 r_2 + \omega_0^2 r_2 = ( g / v_0) \partial_t r_1 + g^2 (r_3 \partial_t r_1 - r_1 \partial_t r_3) / v_0^3~,
\end{eqnarray}
where $\omega_0^2 = ( v_0 / 2 m ) \nabla^* \cdot ( \partial_0 \textbf{p} + \nabla \times \textbf{p} ) - g^2 / 2 v_0^2 $ , $\textbf{p} = m \partial_t \textbf{r} $ , $\textbf{r} = \Sigma ( \emph{\textbf{i}}_j r_j )$ .

Accordingly a series of intricate frequencies, $\omega$, of mass particles can be derived from the above equations in the electric-like field. In considering of more force terms, we may obtain more complicated equations than the above, and find some more elaborate results about the Stark-like effect for mass particles.

It is easy to think out that the Larmor-like frequencies may be shifted in the energy's unsteady states, $\nabla w_0 \neq 0$. The Sark-like effects will be affected by many force terms, and some other factors regarding the electric-like field and electromagnetic field etc.

\subsection{Pockels-like effect for mass particles}

In the gravitational field and electromagnetic field, the definition of force density can conclude some equations, which can be used to explain the Pockels-like effect for electric-like component of gravitational field, which can cause the refractive index to vary approximately in proportion to the electric-like strength.

In case the physical system is on one complex case, and has to consider the effect of $w_0$ and $\textbf{w}$ simultaneously. When $(\textbf{E}/v_0 + \textbf{B}) = 0$, $\textbf{X}_0 = 0$, $\textbf{b} = 0$, $\textbf{g}^* \times \textbf{w}/v_0^2 \approx 0$, and $w_0 \textbf{g}^* / v_0^2 \approx 0$, Eq.(17) will be reduced to
\begin{eqnarray}
- 2 \textbf{f} \approx && \nabla^* w_0 + \partial_0 \textbf{w} + \nabla^* \times \textbf{w}
\nonumber\\
\approx && \nabla^* ( x_0 z_0  + \textbf{x} \cdot \textbf{z} ) + \partial_0 ( x_0 \textbf{z}  + z_0 \textbf{x}) + \nabla^* \times ( x_0 \textbf{z} + z_0 \textbf{x} )
\nonumber\\
\approx && z_0 ( \nabla^* x_0 ) + ( \partial_0 x_0 + \nabla x_0 ) \times \textbf{z}
~,
\end{eqnarray}
where $z_0 = \partial_0 p_0 + \nabla \cdot \textbf{p}$, and $\textbf{z} = \partial_0 \textbf{p} + \nabla p_0 + \nabla \times \textbf{p}$ .

When $\textbf{f} = 0$, the above can be reduced further as follows,
\begin{eqnarray}
0 = z_0 ( - \textbf{a}_E / v_0 + \partial_0 \textbf{x} ) + \textbf{a}_E \times \textbf{z} ~,
\end{eqnarray}
and then
\begin{eqnarray}
0 = ( - \textbf{a}_E / v_0 + \partial_0 \textbf{x} ) + ( 1 / z_0 ) ( \textbf{a}_E \times \textbf{z} ) ~, \nonumber
\end{eqnarray}
or
\begin{eqnarray}
\textbf{a}_E = \partial_t \textbf{x} + \sigma ( \textbf{a}_E \times \textbf{z} ) ~,
\end{eqnarray}
where $\sigma = v_0 / z_0 $ , $\textbf{a}_E = \partial_0 \textbf{x} + \nabla \circ x_0$ , $\textbf{a}_B = \nabla \times \textbf{x}$ . For a tiny volume-distribution region of the optical medium, $\textbf{p}$, $\textbf{a}_E$, $\textbf{a}_B$, and $\sigma$ are all mean values.

Further
\begin{eqnarray}
\partial_t \textbf{a}_E = \partial_t^2 \textbf{x} + \sigma \partial_t ( \textbf{a}_E \times \textbf{z} )~.
\end{eqnarray}

Considering the following relation obtained from the definition equations, $ \mathbb{A} = \lozenge \circ \mathbb{X} $ and $ \mathbb{X} = \lozenge^* \circ \mathbb{Y} $ ,
\begin{eqnarray}
\partial_t \textbf{a}_E = v_0^2 \nabla \times \textbf{a}_B = - v_0^2 \nabla^2 \textbf{x} ~,
\end{eqnarray}
Eq.(147) can be rewritten as
\begin{eqnarray}
v_0^2 \nabla^2 \textbf{x} = \partial_t^2 \textbf{x} + \sigma \partial_t \left\{ ( - v_0^2 \nabla^2 \textbf{x} ) \times \textbf{z} \right\}  ~,
\end{eqnarray}
where $\mathbb{Y}$ is an octonion.

Setting the linearly polarized light transmit along the z-axis in the coordinate system $(r_0, r_1, r_2, r_3)$, the above can be reduced to the equation for electro-optic effect,
\begin{eqnarray}
&& v_0^2 \partial_3^2 x_1 - \partial_t^2 x_1 + M_g z_3  \partial_t ( \partial_3^2 x_2 ) = 0~,
\\
&& v_0^2 \partial_3^2 x_2 - \partial_t^2 x_2 - M_g z_3  \partial_t ( \partial_3^2 x_1 ) = 0~,
\end{eqnarray}
where $z_3$ is the z-axis component of $\textbf{z}$ , $M_g = - v_0^2 \sigma$ .

According to the form of above equations, we can suppose that one solution is
\begin{eqnarray}
&&  x_1 = x cos ( \alpha t - \beta r_3 ) cos ( \gamma t )~,
\\
&&  x_2 = x cos ( \alpha t - \beta r_3 ) sin ( \gamma t )~,
\end{eqnarray}
and making the substitution we find
\begin{eqnarray}
&&  0 = \alpha^2 + \gamma^2 - \beta^2 ( v_0^2  + M_g z_3 \gamma )~,
\\
&&  0 = 2 \gamma - M_g \beta^2 z_3~,
\end{eqnarray}
and then
\begin{eqnarray}
&& \gamma = M_g \beta^2 z_3 / 2~,
\\
&& v^2 = ( \alpha / \beta)^2 =  v_0^2  + M_g z_3 ( 1 - \gamma / 2)~,
\end{eqnarray}
that is,
\begin{eqnarray}
v^2 = v_0^2  + M_g z_3 ( 1 - M_g \lambda^2 z_3 / 8 \pi n^2 )~,
\end{eqnarray}
where $v$ is the speed of light in the medium. $\beta \lambda = 2 \pi n$ , and $\alpha \lambda = 2 \pi v$ .

And then the total angle, $\theta$, of rotation of the light beam will evidently be
\begin{eqnarray}
\theta = \gamma D_g / v = 2 \pi^2 M_g z_3 n^3 D_g / v_0 \lambda_0^2~,
\end{eqnarray}
where $n$ is the refractive index, $\lambda$ is the wavelength, and $D_g$ is the length of substance. $ \beta \lambda_0 = 2 \pi n $ , $ \alpha \lambda_0 = 2 \pi v $ , $ \lambda = \lambda_0 / n $ , and $v = v_0 / n $ . $\lambda_0$ and $v_0$ are the wavelength and the speed of light in the vacuum respectively.

Combining with the dispersion formula, $ d \beta / d \alpha = ( n / v ) \left\{ n - \lambda ( d n / d \lambda ) \right\}$ , the rotation angle of linearly polarized light will be,
\begin{eqnarray}
\theta = M_g' (n / \lambda_0 )^2 \left\{ n - \lambda ( d n / d \lambda ) \right\} z_3 D_g ~,
\end{eqnarray}
where $M_g'$ is a constant coefficient.

Considering the condition formula $\textbf{z} = k' \textbf{j}$ and the Ohm-like law, $\textbf{j} = \sigma' \textbf{g}$ , we obtain when $\partial_t g_3 \approx 0 $
\begin{eqnarray}
z_3 \approx k' ( \partial_t \sigma' ) g_3 ~,
\end{eqnarray}
and then the angle $\theta$ can be written as follows
\begin{eqnarray}
\theta = M_g'' (n / \lambda_0 )^2 \left\{ n - \lambda ( d n / d \lambda ) \right\} g_3 D_g ~,
\end{eqnarray}
where $M_g''$, $k'$, and $\sigma'$ are coefficients. $g_3$ is the z-axis component of $\textbf{g}$ . $\textbf{j}$ is the linear momentum density.

When $\textbf{x}$ and $\textbf{z}$ in Eq.(149) are both 3-dimensional quantities, the obtained rotation angle of linearly polarized light will be more complex than the above.

This inference can be considered as one kind of theoretical explanations for the rotation of linearly polarization light, except for that in the magnetic-field. And that the angle $\theta$ can be affected by many force terms in the electric-like field. The Pockels-like effect is linear in the applied electric-like field, while the Kerr-like effect is quadratic in the electric-like field which should be considered the force terms regarding the $(\textbf{g}/v_0)$ .

\subsection{Self-focusing in the electric-like field}

In the equation for the optic effects in the electric-like field, the transmitted direction of incident ray may be paralleled to that of the applied electric-like field. When these two directions are perpendicular to each other, we will obtain the equations for the double refractions as well as the self-focusing phenomena in the electric-like field.

Setting the linearly polarized light transmit along the z-axis in the coordinate system $(r_0, r_1, r_2, r_3)$, and its polarized direction is along the x-axis, Eq.(149) can be reduced to the equation which is similar to that for Pockels-like effect,
\begin{eqnarray}
&& v_0^2 \partial_3^2 x_1 - \partial_t^2 x_1 + M_g z_2  \partial_t ( \partial_3^2 x_3 ) = 0~,
\\
&& v_0^2 \partial_3^2 x_3 - \partial_t^2 x_3 - M_g z_2  \partial_t ( \partial_3^2 x_1 ) = 0~,
\end{eqnarray}
where $z_2$ is the y-axis component of $\textbf{z}$ , $M_g = - v_0^2 \sigma$ .

According to the form of above equations, we can suppose that one solution is
\begin{eqnarray}
&&  x_1 = x cos ( \alpha t - \beta r_3 ) cos ( \gamma t )~,
\\
&&  x_3 = x cos ( \alpha t - \beta r_3 ) sin ( \gamma t )~,
\end{eqnarray}
and making the substitution we find
\begin{eqnarray}
&&  0 = \alpha^2 + \gamma^2 - \beta^2 ( v_0^2  + M_g z_2 \gamma )~,
\\
&&  0 = 2 \gamma - M_g \beta^2 z_2~,
\end{eqnarray}
and then
\begin{eqnarray}
&& \gamma = M_g \beta^2 z_2 / 2~,
\\
&& v^2 = ( \alpha / \beta)^2 =  v_0^2  + M_g z_2 ( 1 - \gamma / 2)~,
\end{eqnarray}
that is,
\begin{eqnarray}
v^2 = v_0^2  + M_g z_2 ( 1 - M_g \lambda^2 z_2 / 8 \pi n^2 )~,
\end{eqnarray}
where $v$ is the speed of light in the medium. $\beta \lambda = 2 \pi n$ , and $\alpha \lambda = 2 \pi v$ .

And then the total angle, $\theta$, of rotation of the light beam will evidently be
\begin{eqnarray}
\theta = \gamma D_g / v = 2 \pi^2 M_g z_2 n^3 D_g / v_0 \lambda_0^2~,
\end{eqnarray}
where $n$ is the refractive index, $\lambda$ is the wavelength, and $D_g$ is the length of substance. $ \beta \lambda_0 = 2 \pi n $ , $ \alpha \lambda_0 = 2 \pi v $ , $ \lambda = \lambda_0 / n $ , and $v = v_0 / n $ . $\lambda_0$ and $v_0$ are the wavelength and the speed of light in the vacuum respectively.

Combining with the dispersion formula, $ d \beta / d \alpha = ( n / v ) \left\{ n - \lambda ( d n / d \lambda ) \right\}$ , the rotation angle of linearly polarized light will be,
\begin{eqnarray}
\theta = M_g' (n / \lambda_0 )^2 \left\{ n - \lambda ( d n / d \lambda ) \right\} z_2 D_g ~,
\end{eqnarray}
where $M_g'$ is a constant coefficient.

Considering the condition formula $\textbf{z} = k' \textbf{j}$ and the Ohm-like law, $\textbf{j} = \sigma' \textbf{g}$ , we obtain when $\partial_t g_2 \approx 0 $
\begin{eqnarray}
z_2 \approx k' ( \partial_t \sigma' ) g_2 ~,
\end{eqnarray}
and then the angle $\theta$ can be written as follows
\begin{eqnarray}
\theta = M_g'' (n / \lambda_0 )^2 \left\{ n - \lambda ( d n / d \lambda ) \right\} g_2 D_g ~,
\end{eqnarray}
where $M_g''$, $k'$, and $\sigma'$ are coefficients. $g_2$ is the y-axis component of $\textbf{g}$ . $\textbf{j}$ is the linear momentum density.

The incident ray may possess two polarized components. One is along the x-axis and the locus is the hyperbolic curve with the rotation angle $\theta$ in Eq.(175), while the other along the y-axis and its direction keeps the same. That is the double refractions phenomena in the electric-like field. In the interstellar matter, the polarized component parallel to the applied field keeps its transmitted orientation, while the polarized component perpendicular to the applied field changes its transmitted orientation. When the incident ray is transmitted along the z-axis, there will exist the refracted ray and the double refractions, such as the Pockels-like effect. Specifically, the applied electric-like field is symmetric with respect to the z-axis, there will be the self-focusing phenomena.

\begin{table}[h]
\caption{\label{tab:table1}The comparison of the analogous optic effects in the electric-like and magnetic-like fields.}
\begin{ruledtabular}
\begin{tabular}{lll}
magnetic-like~field                                 &  electric-like~field                             &  description                 \\
\hline
Larmor-like precession                              &  (analogue)                                      &  mass particle               \\
Zeeman-like effect                                  &  Stark-like effect                               &  mass particle               \\
Faraday-like effect (first order)                   &  longitudinal Pockels-like effect                &  parallel orientations       \\
(analogue)                                          &  transversal Pockels-like effect                 &  perpendicular orientations  \\
Faraday-like effect (second order)                  &  longitudinal Kerr-like effect                   &  parallel orientations       \\
Voigt-like and Cotton-Mouton-like effects           &  transversal Kerr-like effect                    &  perpendicular orientations  \\
reflected Kerr-like effect                          &  (analogue)                                      &  reflected light             \\
self-focusing                                       &  self-focusing                                   &  axial symmetric             \\
\end{tabular}
\end{ruledtabular}
\end{table}

\section{CONCLUSIONS}

In the electromagnetic field and gravitational field described with the algebra of octonions, the magneto-optic effect and electro-optic effect are derived from the same force definition formula. In the octonion space, the Zeeman effect is deduced from the terms $(\partial_0 \textbf{w} + \nabla^* \times \textbf{w} + k_{eg} \textbf{B}^* \times \textbf{W}/v_0)$, Faraday effect from the term $\nabla^* w_0$ , Stark effect from the terms $(\partial_0 \textbf{w} + \nabla^* \times \textbf{w} + k_{eg} \textbf{E}^* \times \textbf{W}/v_0^2)$, and Pockels effect from the terms $(\nabla^* w_0 + \partial_0 \textbf{w} + \nabla^* \times \textbf{w})$. Moreover the rotation of linearly polarized light is affected by the force terms regarding the gravitational field also.

The above theoretical explanation is different to the classical index ellipsoid approach. This work indicates that the above inferences will be impacted by not only the refractive index and field energy but also the field strength and other factors. And that the gravitational strength $\textbf{g}$ and $\textbf{b}$ will affected the orientational rotation of linearly polarized light, although the rotation affection may be quite tiny. However in the universe space the rotation angle of the gravitational affection may be distinct enough to be detected due to the long distance transmission.

It should be noted that the study for the optic effects impacted by the force terms examined only some simple cases under the octonion force definition came from the torque and energy. Despite its preliminary character, this study can clearly indicate that the magneto-optic effect and electro-optic effect both can be derived from the same one force definition in the electromagnetic field and gravitational field. For the future studies, the research will focus on the theoretical explanations about the birefringent phenomenon in the optical medium as well as the experimental validations of the rotation of linearly polarized light in the gravitational field.

\begin{acknowledgments}
This project was supported partially by the National Natural Science Foundation of China under grant number 60677039.
\end{acknowledgments}


\begin{references}

\bibitem{larmor}
      Larmor, J.,
      {\it Aether and matter},
      (The Cambridge University Press, Cambridge, 1900).

\bibitem{fitzgerald}
      Fitzgerald, G. F.,
      {\it The scientific writings of the late George Francis Fitzgerald: collected and edited with a historical introduction},
      (ed. J. Larmor),
      (Hodges, Figgis \& Co., Ltd., Dublin, 1902).

\bibitem{zeeman}
      Zeeman, P.,
      {\it Researches in magneto-optics},
      (Macmillan and Co., Ltd., London, 1913).

\bibitem{faraday}
     Whittaker, E. T.,
     {\it A history of the theories of aether and electricity: from the age of Descartes to the close of the nineteenth century},
     (Longmans, Green and Co., London, 1910).

\bibitem{stark}
      Bini, D., C. Cherubini, and B. Mashhoon,
      ``Vacuum C-metric and the Gravitational Stark Effect",
      {\it Physical Review D\/},
      Vol.~70, No.~4, 044020, 7 pages, 2004.

\bibitem{pockels}
      Rao, K. V., and T. S. Narasimhamurty,
      ``Pockels' Effect in $\alpha$-quartz",
      {\it Journal of Modern Optics\/},
      Vol.~19, No.~4, 319--325, 1972.

\bibitem{kerr}
      Kerr, P.,
      {\it An elementary treatise on rational mechanics},
      (William Hamilton, Glasgow, 1866).

\bibitem{maccullagh}
      MacCullagh, J. H.,
      {\it The collected works of James MacCullagh},
      (ed. J. H. Jellett and S. Haughton)
      (Longmans, Green \& Co., London, 1880).

\bibitem{maxwell}
      Maxwell,~J.~C.,
      {\it A Treatise on Electricity and Magnetism},
      Dover Publications Inc., New York, 1954.

\bibitem{rowlland1}
      Rowlland, H. A.,
      {\it The physical papers of Henry Augustus Rowlland},
      (The Johns Hopkins Press, Baltimore, 1902).

\bibitem{barrett}
      Barrett, T. W.,
      ``Changes in the refractive index ellipsoid isotropies, symmetric anisotropies, and depolarization ratios of potassium hyaluronate solutions as a function of pH",
      {\it Biopolymers\/},
      Vol.~18, No.~2, 351--358, 1979.

\bibitem{nawaz}
      Nawaz, M., W. A. Farooq, and J.-P. Connerade,
      ``The influence on the Paschen-Back effect on magneto-optical rotation spectra",
      {\it Journal of Physics B: Atomic, Molecular, and Optical Physics\/},
      Vol.~25, No.~15, 3283--3294, 1992.

\bibitem{cayley}
      Cayley, A.,
      {\it The collected mathematical papers of Arthur Cayley},
      (The Cambridge University Press, Cambridge, 1889).

\bibitem{hamilton}
      Hamilton, W. R.,
      {\it Elements of Quaternions},
      (Longmans, Green \& Co., London, 1866).

\bibitem{weng1}
      Weng,~Z.-H.,
      ``Electromagnetic Forces on Charged Particles",
      in {\it Progress in Electromagnetics Research Symposium 2009 in Moscow Proceedings}, 361--363,
      Moscow Russia, August 18-21, 2009.

\bibitem{weng2}
      Weng,~Z.-H.,
      ``Octonionic Electromagnetic and Gravitational Interactions and Dark Matter",
      (arXiv:physics/0612102)

\bibitem{newton}
     Newton, I.,
     {\it The Mathematical Principles of Natural Philosophy},
     {trans. A. Motte},
     (Dawsons of Pall Mall, London, 1968).

\bibitem{lorentz}
     Lorentz, H. A.,
     {\it The Theory of Electrons},
     (Dover Publications Inc., New York, 1952).

\bibitem{uhlenbeck}
      Uhlenbeck, G. E., and S. Goudsmit,
      ``Spinning Electrons and the Structure of Spectra",
      {\it Nature}, Vol.~117, No.~2938, 264--265, 1926.

\bibitem{rowlland2}
      Rowlland, H. A.,
      ``On the new theory of magnetic attractions, and the magnetic rotation of polarized light",
      {\it Philosophical Magazine Series 5\/},
      Vol.~11, No.~68, 254--261, 1881.

\bibitem{heaviside}
      Heaviside, O.,
      ``A Gravitational and Electromagnetic Analogy",
      {\it The Electrician}, Vol.~31, 281--282, 359, 1893.



\end{references}
\end{document}